\documentclass[pra, notitlepage, superscriptaddress,  reprint]{revtex4-1}
\usepackage{amssymb}
\usepackage{amsmath}
\usepackage{braket}
\usepackage{graphicx}
\usepackage[usenames,dvipsnames]{color}
\usepackage{tensor}
\usepackage{empheq}
\usepackage{capt-of}
\usepackage[normalem]{ulem} 
\usepackage{soul} %

\definecolor{myorange}{RGB}{199.24, 87.48, 47.80}
\usepackage[colorlinks,bookmarks=false,citecolor=NavyBlue,linkcolor=OliveGreen,urlcolor=blue]{hyperref}
\newcommand{\be}{\begin{equation}}
\newcommand{\ee}{\end{equation}}
\newcommand{\ba}{\begin{aligned}}
\newcommand{\ea}{\end{aligned}}
\newcommand{\bw}{\begin{widetext}}
\newcommand{\ew}{\end{widetext}}

\newcommand{\tr}[2]{\mathrm{tr}_{#1}(#2)}

\usepackage{comment}
\newcommand{\lv}{{\boldsymbol \lambda}}

\newcommand{\titleinfo}{Non-equilibrium spin transport in integrable spin chains:\\ 
persistent currents and emergence of magnetic domains}

\def\epp{\: .}
\def\epc{\: ,}

\def\smax{\hat{s}}
\def\id{\mathbf{1}}

\def\doi{http://dx.doi.org/}

\begin{document}

\title{\titleinfo}

\author{Andrea De Luca}
\affiliation{The Rudolf Peierls Centre for Theoretical Physics, Oxford University, Oxford, OX1 3NP, United Kingdom}
\author{Mario Collura}
\affiliation{The Rudolf Peierls Centre for Theoretical Physics, Oxford University, Oxford, OX1 3NP, United Kingdom}
\author{Jacopo De Nardis}
\affiliation{D\'epartement de Physique, \'Ecole Normale Sup\'erieure \\ PSL Research
University, CNRS, 24 rue Lhomond, 75005 Paris, France}

\begin{abstract}

We construct exact steady states of unitary non-equilibrium time evolution in the gapless XXZ spin-1/2 chain where 
integrability preserves ballistic spin transport at long-times.
We characterize the quasi-local conserved quantities responsible for this feature
and introduce a computationally effective way to evaluate their expectation values 
on generic matrix product initial states. We employ this approach to reproduce 
the long-time limit of local observables in all quantum
quenches which explicitly break particle-hole or time-reversal symmetry. 
We focus on a class of initial states supporting persistent spin currents
and our predictions remarkably agree with numerical simulations at long times. 
Furthermore, we propose a protocol for this model where
interactions, even when antiferromagnetic, are responsible for the 
unbounded growth of a macroscopic magnetic domain. 
\end{abstract}

\maketitle

\paragraph*{Introduction. ---}

One-dimensional many-body physics displays a rich landscape of exactly solvable models, which offers 
a perfect outpost to study the role of interaction in quantum mechanics. As a remarkable example, integrable spin chains 
have provided in the past decades, numerous insights on the equilibrium properties of quantum magnets, 
then confirmed in several experiments \cite{expXXZ}. 
On the other hand, the study of non-equilibrium properties has a much shorter history, as only recently 
isolated quantum systems have been successfully prepared in experiments and studied under their unitary time evolution 
\cite{exp_unitary_time_evolution}. In these settings, it was possible to observe with high precision transport processes in many-body 1d systems \cite{exp_transport} and relaxation to steady states after a non-equilibrium time evolution \cite{exp}.

From the theoretical point of view, the focus has been on understanding how a unitarily evolving system 
can relax to a stationary state \cite{EF:review} and, if this is the case, 
how to predict the stationary values of local observables. {  The XXZ spin-$1/2$ chain defined by the Hamiltonian \eqref{Eq:Hamiltonian} represents a perfect setting where to understand the role of interactions and local symmetries in the non-equilibrium time evolution \cite{quench-xxz-fagotti,quench-xxz,Piroli_XXZ,IlievskiPRLQL,IlievskiJSTAT}. }
In particular, in one dimension, two main non-equilibrium protocols have been considered: a) the so called quantum quench, 
when starting from a pure state $| \Psi_0 \rangle$, the system is let unitarily evolve under its Hamiltonian $| \Psi_0(t) \rangle = e^{- i H t } | \Psi_0 \rangle$~\cite{quench-xxz} and b) 
the open Markovian case, when the edges of the chain are kept in contact with two spin baths \cite{Prosen_Open}
(a similar protocol has also recently been realized in experiments~\cite{Hirobe}). 
The study of the case a) has led to a neat formulation \cite{IlievskiPRLQL,IlievskiJSTAT}  
of the Generalized Gibbs Ensemble (GGE)~ \cite{GGEcit}: 
the steady state is locally described by a density matrix of Gibbs-like form 
$\propto e^{- \sum_j \beta_j \boldsymbol{Q}_j}$, with 
$\{ \boldsymbol{Q}_j\}$ a set of local and quasi-local conserved quantities, 
namely extensive operators commuting with the global Hamiltonian $[\boldsymbol{H}, \boldsymbol{Q}_j] = 0$. 
The $\beta$'s are fixed requesting that the stationary expectation values of all the conserved charges
coincide with the initial ones. This statement 
extends to any interacting model once an appropriate set $\{ \boldsymbol{Q}_j\}$ has been identified.

The protocol b) has shown how steady states can exhibit a genuine non-equilibrium behavior and 
persistent spin currents. At the origin of this peculiar behavior, lies the existence of a set of 
of previously unknown quasi-local conserved operators $\{ \boldsymbol{Z}_n\}$ which preserve ``ballistic'' spin transport \cite{Prosen_Open_QL}.
This set of conserved operators is characteristic of the gapless phase and is responsible for the existence of a 
finite Drude weight~\cite{Prosen_Open_QL,PereiraDrude, ProsenDrude}, which has attracted large 
interest in the recent years \cite{Drude_XXZ_preProsen}.

However, it was unclear if such quasi-local operators $\{ \boldsymbol{Z}_n\}$ could be included into a GGE-like steady state resulting from the long-time dynamics of a close system. 
If this is the case, one can choose initial states $| \Psi_0 \rangle$, which, by pure unitary time evolution,
lead to a persistent current in the stationary state, namely 
$\lim_{t \to \infty} \langle   \Psi_0(t)|\boldsymbol{J}_i |\Psi_0(t) \rangle \propto 
\text{Tr} \bigl( e^{- \sum_j \beta_j \boldsymbol{Q}_j - \sum_j \gamma_j \boldsymbol{Z}_j} \boldsymbol{J}_i  \bigr)\neq 0
$. Here current operators are defined by continuity equations, e.g. for the local magnetizations, 
$d \boldsymbol{s_i^z}/dt = \imath [\boldsymbol{H}, \boldsymbol{s_i^z}] = \boldsymbol{J}_{i-1}-\boldsymbol{J}_{i}$.

In this work, we solve this open issue. We show that 
the unitary evolution of a generic state $| \Psi _0 \rangle $ with the gapless XXZ Hamiltonian,
leads to a GGE steady state which includes the set $\{ \boldsymbol{Z}_n\}$. 
Ignoring these quantities in the GGE produces significative discrepancies 
with the large-time limit of local observables, as was observed 
in a specific example \cite{Prosen_flux_quench}. Moreover,
we present an efficient way to handle such charges in practice, 
which is applicable to any initial state expressed as a matrix product state.  
We obtain a class of non-equilibrium time evolutions exhibiting persistent spin transport at large times, beyond
the linear response regime described by the equilibrium Drude weight
(accessible isotopicinstead with local quantum quenches \cite{Prosen_Prosal}). Moreover,
thanks to our construction, non-trivial 
steady state can be engineered by joining two initial states with different spin-transport properties. 
In particular, we propose an experimentally feasible protocol 
to create a macroscopic magnetic domain inside the antiferromagnetic 
phase of the XXZ chain \eqref{Eq:Hamiltonian}. 
 \paragraph*{The model and classification of steady states. ---}
We consider the XXZ spin-$1/2$ chain in the gapless phase $|\Delta|< 1$ 
\be
{\boldsymbol H}= \sum_{i=-\frac{L}{2}+1}^{\frac{L}{2}}\Bigl[{\boldsymbol s}^{x}_i {\boldsymbol s}^{x}_{i+1}+  {\boldsymbol s}^{y}_i {\boldsymbol s}^{y}_{i+1}+\Delta \left( {\boldsymbol s}^{z}_i {\boldsymbol s}^{z}_{i +1}-\frac{1}{4}\right)\Bigr],
\label{Eq:Hamiltonian}
\ee
(with $\{{\boldsymbol s}^{\alpha}_i\}$ indicating the spin-$\frac{1}{2}$ operators) 
and parametrise the anisotropy as $\Delta=\cos\gamma$. The model is solvable by Bethe ansatz: every eigenstate $\ket{\{\lambda_i\}}$ is labeled by a set of $N$ complex ``rapidities''  $\{\lambda_k\}_{k=1}^N$ fulfilling the Bethe equations \cite{supp_matt}. 
In the thermodynamic limit $L \to \infty$, rapidities are arranged
in strings, composed of $n_j$ rapidities, sharing the same real part $\lambda_\alpha^{(j)}$ and equispaced imaginary parts, with gaps of size $\imath \gamma$. These solutions can be interpret as bound states of magnons with different lengths $\{n_j\}_{j=1}^{N_s}$ and center of mass momenta $\{ \lambda_\alpha^{(j)} \}_{j=1}^{N_s}$ with $\alpha$ labeling all bound states with the same length. 
Whenever $\gamma/\pi$ is a rational number, the total number of string types is finite.
In the following, we focus on the simplest case $\gamma = \pi/\ell$, where  
one has exactly $\ell$ bound-state types.

In the thermodynamic limit the string momenta $\{\lambda_\alpha^{(j)}\}_\alpha$ become dense on the real line and for each string type $j$,
we can introduce the corresponding density distribution 
$L \rho_j(\lambda_\alpha^{(j)}) = (\lambda_{\alpha+1}^{(j)}-\lambda_\alpha^{(j)})^{-1}$. 
We define then a macroscopic eigenstate $\ket{\rho}$~\cite{Korepinbook} as a set of particles distributions 
$\{\rho_j(\lambda)\}_{j=1}^{\ell}$, one for each string type. 
Then, the expectation value of any extensive conserved charge $\boldsymbol{Q} = \sum_i \boldsymbol{q}_i$, 
with $\boldsymbol{q}_i$ the charge density, is expressed in the thermodynamic limit only in terms of the 
distribution of particles and its single-particle eigenvalues $q_j(\lambda)$
\begin{equation}\label{exp_charges_rho}
\langle \rho  | \boldsymbol{Q} | \rho \rangle =L \sum_{j=1}^{\ell}  \int_{-\infty}^{+\infty}  d\lambda \ \rho_j(\lambda) q_{j}  (\lambda)\,.
\end{equation}
In particular, the total number of particles, i.e. the number 
of down spins in the system $\boldsymbol{Q} = L/2 - \sum_{i}\boldsymbol{s}^{z}_i$, corresponds to the case $q_j(\lambda) = n_j$ 
and fixes the normalization of the densities $\{\rho_j(\lambda)\}_{j=1}^{\ell}$. Instead, the energy density 
$e = \frac{1}{L}\langle \rho  | \boldsymbol{H} | \rho \rangle $  
is obtained for $q_j = - J\pi  \sin(\gamma)  a_j  (\lambda)$, with $a_j(\lambda)$ the scattering kernel given in (S28) of \cite{supp_matt}.

\begin{figure}
 \includegraphics[width=0.35\textwidth]{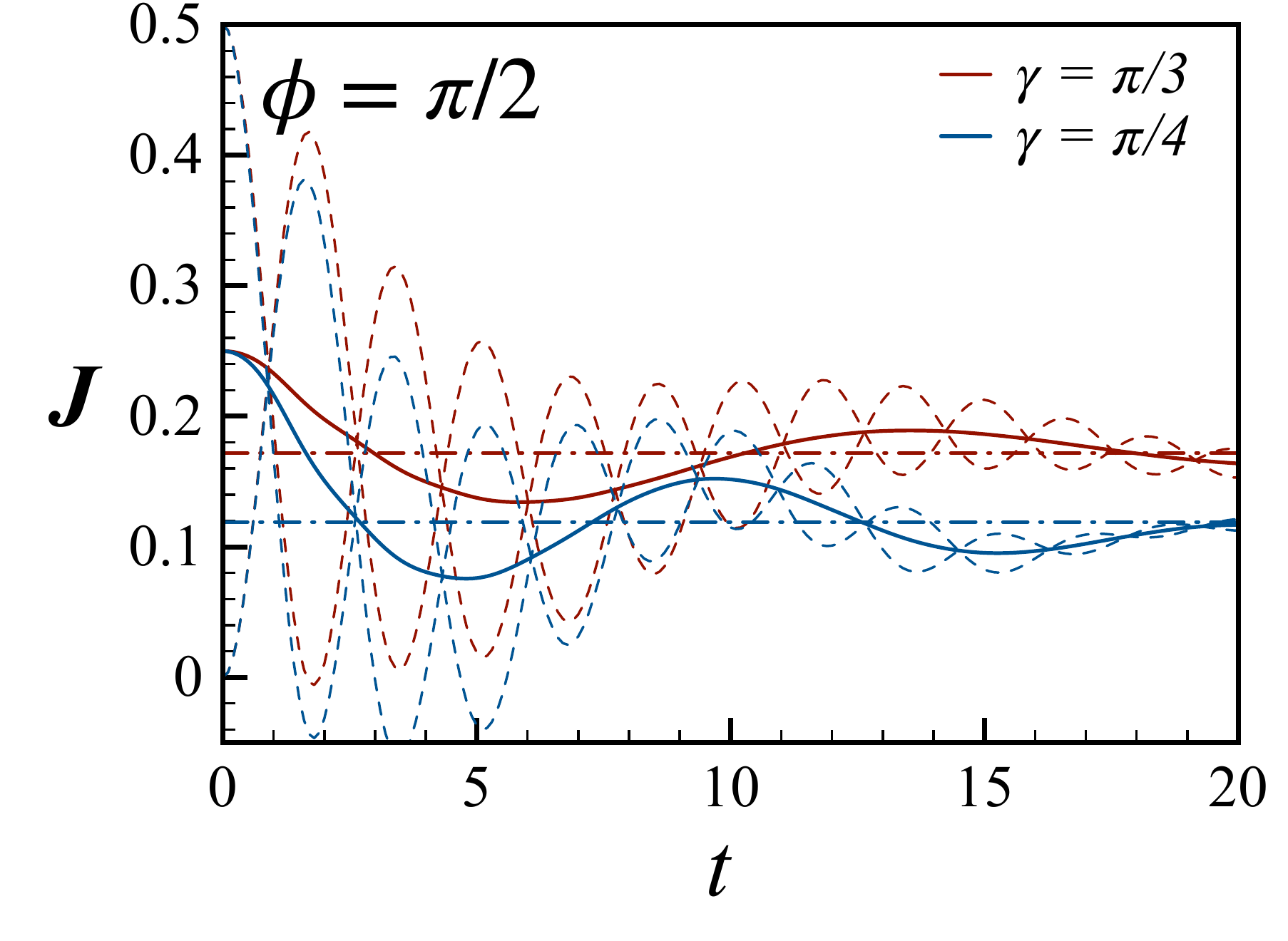}
 \caption{\label{fig:time_ev_spin_dimer}
 Time evolution of the expectation value of the spin current operator $\boldsymbol{J}_i= \bigl( {\boldsymbol s}^{x}_i {\boldsymbol s}^{y}_{i+1}- 
{\boldsymbol s}^{y}_i {\boldsymbol s}^{x}_{i+1}  \bigr)$
 obtained by iTEBD simulations with Trotter time-step $dt = 0.01$ 
 and maximum auxiliary dimension $\chi = 1024$, starting from the dimer state 
 \eqref{dimer_state} with $\phi = \pi/2$ for two different values of $\Delta=\cos \gamma$. 
 The horizontal dot-dashed lines represent the steady state predictions given by the complete GGE \eqref{GGE_odd}. 
 The dashed lines represent the time evolution of the current operator evaluated on odd and even sites while the full line their average. 
 }
\end{figure}

Analogously to representing the Gibbs trace with a single expectation value at thermal equilibrium \cite{ETH}, it is useful to associate to the canonical GGE density matrix 
$\propto  e^{- \sum_j \beta_j \boldsymbol{Q}_j}$, 
a corresponding microcanonical formulation, where expectation values are computed 
on a single macroscopic eigenstate  \cite{QA}
\begin{equation}
\label{canomicro}
\frac{\text{Tr} \left( e^{- \sum_j \beta_j \boldsymbol{Q}_j}  \mathcal{O} \right)}{\text{Tr} \left( e^{- \sum_j \beta_j \boldsymbol{Q}_j}   \right)} =  {\langle \rho | \mathcal{O} |\rho \rangle }{}
\end{equation}
for any local operator $\mathcal{O}$. 
The densities $\{ \rho_j\}_{j=1}^{\ell}$ associated to the macroscopic eigentate in \eqref{canomicro}
are fixed matching the expectation values \eqref{exp_charges_rho} 
of the conserved charges. 
A \textit{complete} set of charges $\{\boldsymbol{Q}_j\}$ is such that 
all the densities $\{ \rho_j\}_{j=1}^{\ell}$ are univocally pinned down.
A set symmetric under spin reversal was introduced in \cite{ProsenQLSymmetric}:
in this approach, charges were organized into a finite number of families 
$\{ \boldsymbol{Q}^{(s)}_n\}$ where $s =1/2,1, \ldots,\hat{s}$ 
(with $\hat{s} = \frac{\ell-1}{2}$), in one-to-one correspondence with irreducible representations of the $SU_q(2)$ 
algebra ($q = e^{\imath \gamma}$) with $s$
labeling the spin of the representation. Here, $n \in \mathbb{N}$ indexes all charges 
for a given representation $s$. For instance, the Hamiltonian \eqref{Eq:Hamiltonian} 
coincides with $\boldsymbol{H}= - J \pi \sin(\gamma) \boldsymbol{Q}^{(1/2)}_1$. 
In \cite{IlievskiPRLQL}, it was found an economic way to directly relate 
the densities $\{\rho_j\}_{j=1}^\ell$ to the expectation values of the charges on any state $\ket{\Psi_0}$:
thanks to \eqref{canomicro}, this provides directly the stationary GGE whenever $\ket{\Psi_0}$ is chosen as an initial state.
This goes through the introduction of the generating functions defined as \cite{IlievskiPRLQL,IlievskiJSTAT} 
\begin{equation}\label{generating_even}
X_j(\lambda) = \frac{1}{ L }\sum_{n=1}^\infty \frac{\lambda^{n-1}}{(n-1)!}  {\langle \Psi_0 | \boldsymbol{Q}^{(j/2)}_n | \Psi_0 \rangle }\epp
 \end{equation}
Despite their representation in terms of a complicated power series, these functions can be easily computed on any 
initial state of the form of a matrix product state \cite{quench-xxz-fagotti}. 
 \begin{figure}
 \includegraphics[width=0.35\textwidth,keepaspectratio]{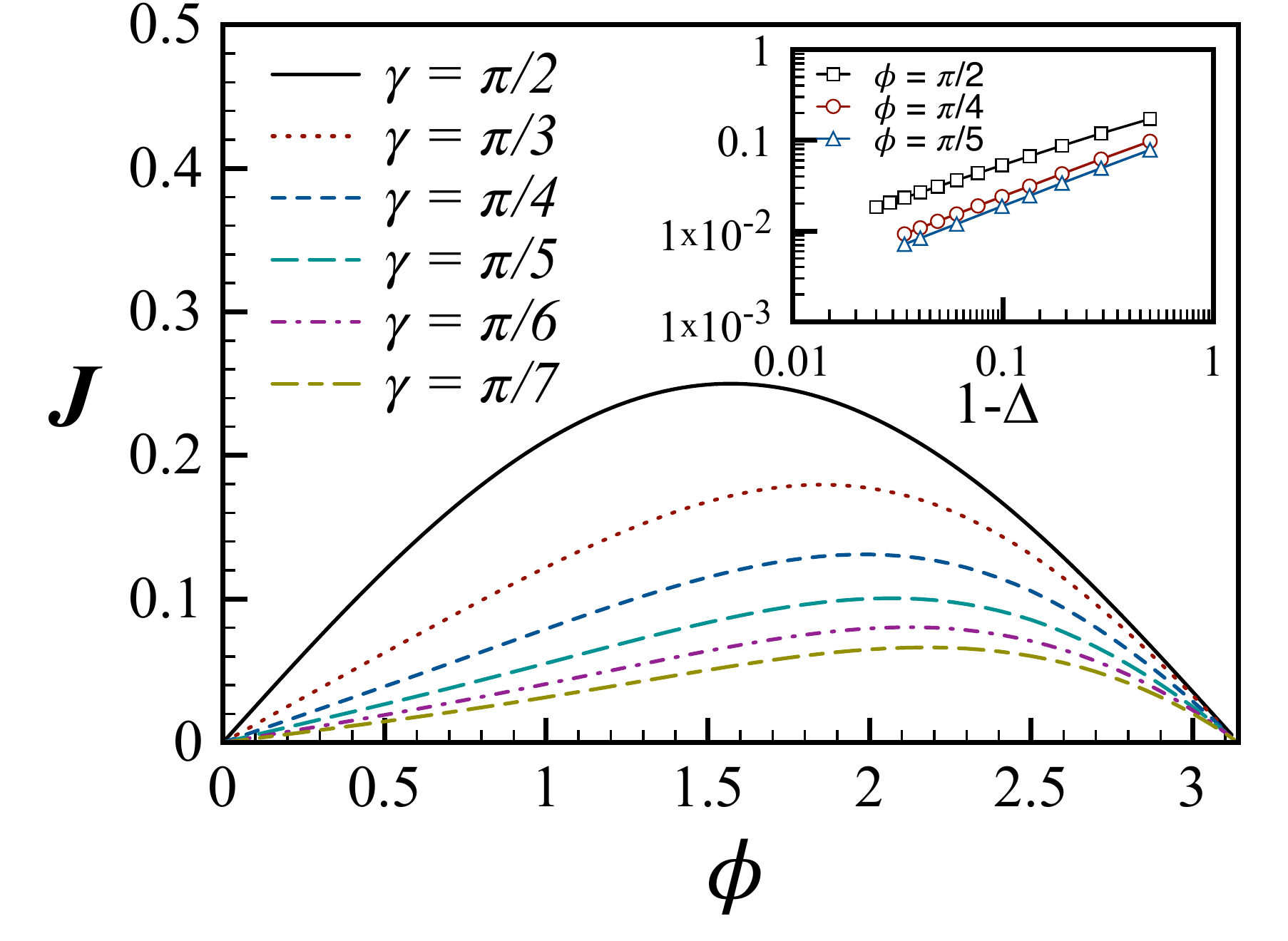}
 \caption{\label{fig:current_dimer}
Steady-state spin current  
after a quench from the Dimer state $| \Psi_0^{(\phi)}\rangle$ \eqref{dimer_state}, as function 
of $\phi$ and for different values of $\Delta = \cos\gamma$. 
Note that the initial current for arbitrary $\Delta$ coincides with 
the stationary current at $\Delta = 0$ ($\gamma = \pi/2$).
Negative values of $\Delta$ can be obtained through the relation: $\langle \boldsymbol{J} \rangle_{\Delta, \phi} = 
\langle \boldsymbol{J} \rangle_{- \Delta, \pi - \phi}$.
In the inset, we show a log-log plot of the stationary current as function of $1-\Delta$ and for three different values of $\phi$.
Data are consistent with a power-law decay $(1-\Delta)^\beta$, with a $\phi$-depedent exponent $\beta(\phi)$. We obtain $\beta(\pi/2) \approx 0.78$,  $\beta(\pi/4) \approx 0.88$ and  $\beta(\pi/5) \approx 0.9$. 
}
\end{figure}
Then, the correspondence between such generating functions and the distributions of roots 
can be deduced through \eqref{exp_charges_rho} in \eqref{generating_even} and reads
\begin{align}\label{GGE_even}
& \rho_j = - X_{j-1} + X_j^{[\pm]} - X_{j+1}\, , \quad j =2, \ldots, \ell -2\,, \nonumber \\
& \rho_1 = X_1^{[\pm]} - X_2\,, \quad  \rho_{\ell - 1} - \rho_\ell = - X_{\ell -2} + X^{[\pm]}_{\ell -1}  \epc
\end{align}
where we used the notation $f^{[\pm]} = f(\lambda +\imath\frac{ \gamma}{2}) +   f(\lambda- \imath\frac{ \gamma}{2})$.

At this point, one immediately realizes that only the difference of the last two distribution is fixed by \eqref{GGE_even}. 
In \cite{IlievskiJSTAT}, the additional constraint was deduced mimicking states describing thermal equilibrium.
However, its validity is found to be restricted to spin-flip invariant initial states 
$\mathcal{S} | \Psi_0 \rangle = \pm | \Psi_0 \rangle$ 
with $\mathcal{S}=\prod_{i} \boldsymbol{s}^x_i $ the spin-flip operator 
on the whole chain (examples are the well-studied Neel and Majumdar-Ghosh states \cite{quench-xxz,Piroli_XXZ}).
We now show how to extend this condition to general quench protocols.

\paragraph*{Generalized Gibbs Ensemble in the gapless regime. ---}
The set of conserved charges included in the GGE \eqref{GGE_even} is not complete in the gapless regime of the XXZ chain.
Here, we provide the remaining set of charges that completely determines any steady state. 
In \cite{PereiraDrude}, it was shown that the key point to get odd conserved quantities in the periodic XXZ spin chain
lies in the representation of $SU_q(2)$ corresponding to maximal spin $s = \smax $.
Indeed, in this case, further irreducible representations of dimension $\ell-1$ 
can be obtained as a function of an additional continuous parameter 
$\alpha$ \cite{supp_matt_representation}. 

Similarly to what is done in \eqref{generating_even}, one can construct 
an $\alpha$-dependent generating function $X_{2\smax, \alpha}(\lambda)$, 
where $\alpha = 0$ reduces to the original $X_{2\smax}(\lambda)$.  
The first difficulty one faces in exploiting, in practice, these operators 
is the illusive richness of new conserved quantities (an infinite family for each value of $\alpha$) 
which contrasts with the limited freedom left in \eqref{GGE_even}, where essentially
only one function remains to be fixed. This apparent paradox can be solved by showing 
that only $X'(\lambda) \equiv \partial_\alpha X_{2\smax,\alpha}(\lambda)|_{\alpha=0}$
has to be added to completely fix all the density distributions.
Up to a multiple of the identity,  $X'(\lambda)$  generates expectation values for a family 
of  conserved quantities $\boldsymbol{Z}_{n}$,
odd under spin flip ($\mathcal{S} \boldsymbol{Z}_{n}\mathcal{S} = - \boldsymbol{Z}_{n}$), i.e.
\begin{equation} \label{gen_xprime}
X'(\lambda) + \mathfrak{H}(\lambda) = \frac{1}{L} \sum_{n=1}^\infty  \frac{\lambda^{n-1}}{(n-1)!}  {\langle \Psi_0 | \boldsymbol{Z}_n | \Psi_0 \rangle }{}  \epc
 \end{equation}
which, as shown in \cite{supp_matt}, can be efficiently computed 
for all initial states $| \Psi_0 \rangle$  expressible as low bond-dimension matrix product states. 
The term $\mathfrak{H}(\lambda) =  \frac{\gamma}{2 \pi \cosh^2( \lambda) }$ 
is simply a multiple of the identity which is added for consistency with all the other $X_j(\lambda)$, all having 
vanishing expectation on the fully polarized state along the $z$-direction.

The second difficulty is to derive a relation between $X'(\lambda)$ and $\{\rho_j(\lambda)\}_{j=1}^{\ell}$, 
giving the missing piece in \eqref{GGE_even}. Being quasi-local, the expectation value of $\boldsymbol{Z}_n$ 
on a many-body state is additive on its particle content, in agreement with \eqref{exp_charges_rho},
but single-particle eigenvalues corresponding to each $\boldsymbol{Z}_n$, 
i.e. the functions $q_j(\lambda)$ in \eqref{exp_charges_rho}, are not accessible with standard methods~\cite{IlievskiPRLQL}.
We follow therefore a different strategy: we consider eigenstates $\boldsymbol{H}$ with a single magnon of rapidity $\tilde \lambda$.
On these states, the expectation value of each $\boldsymbol{Z}_n$ can be analytically computed, 
and gives access in the limit $L \to \infty$
to the required functions $q_j(\tilde \lambda)$~\cite{supp_matt}.

With this information, 
we can complete \eqref{GGE_even} with the additional relation
\begin{align}\label{GGE_odd}
& \rho_\ell =  -  \frac{1}{2} X^{[\pm]}_{\ell -1}   - \frac{1}{ 2 \gamma} \int_{-\gamma/2}^{\gamma/2} dz \ X'^{[z]} \epc
\end{align}
where for a generic function $f$, we introduced the notation $f^{[z]}(\lambda) \equiv f(\lambda + \imath z)$.
When the initial state $| \Psi_0 \rangle $ is invariant under spin-flip inversion we have  $\langle \Psi_0 | \boldsymbol{Z}_n | \Psi_0 \rangle =0$ and $X'(\lambda)  = - \mathfrak{H}(\lambda)$. 
This can be shown to imply the restriction assumed in \cite{IlievskiJSTAT} (see \cite{supp_matt}), 
which is therefore proved to be correct only for spin-flip invariant initial states. 
For all other cases, $X'(\lambda)$ is non-trivial and 
it is responsible for a steady spin current in the system. 
We consider an explicit example in the following section. 
\begin{figure}[t]
\includegraphics[width=0.4\textwidth,keepaspectratio]{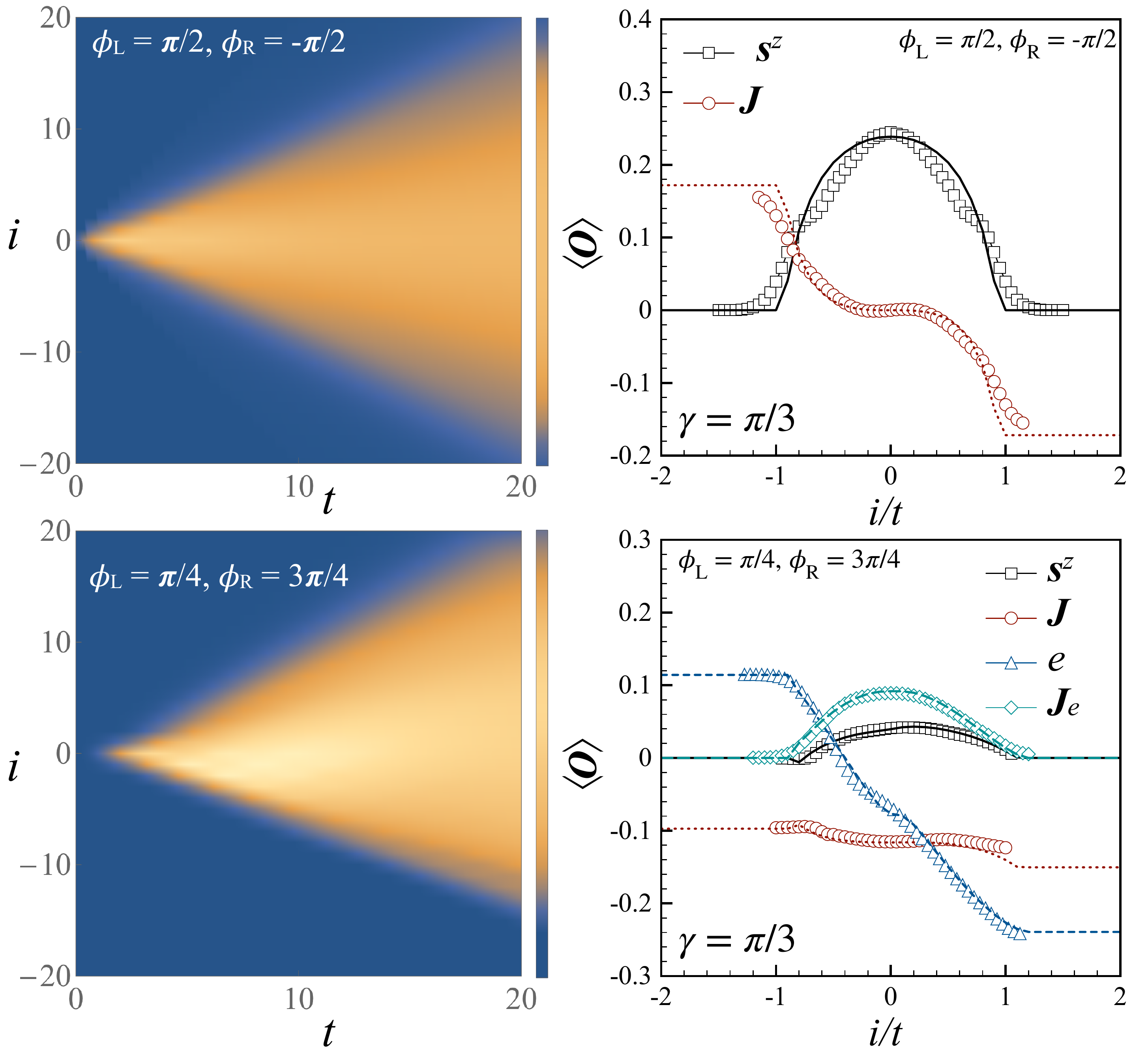}
 \caption{\label{fig:Profiles_time_ev_spin_dimer} Time evolution of local quantities at $\Delta=1/2$ obtained by 
 TEBD simulations from the initial state \eqref{two_dimers}.
 Space and time oscillations in the numerical data were smoothed out by taking the space average 
 over two sites and the time average in $[0,t]$ at fixed rescaled space $i/t$.
(Left) A domain with finite magnetization opens insight the light cone and it approaches the 
rescaled stationary profile (color strips). 
(Right) Expectation value for magnetization and energy densities $\boldsymbol{s}^z_i$, $\boldsymbol{e}_i$ 
and their respective currents $\boldsymbol{J}_i$, $\boldsymbol{J}_{e,i}$
vs rescaled space $i/t$. $\boldsymbol{J}_{e,i}$ is define through the continuity equation
$d \boldsymbol{e_i}/dt = \boldsymbol{J}_{e,i-1}-\boldsymbol{J}_{e, i}$.
The thick lines  correspond to the local steady states prediction for $t \to \infty$. 
}
\end{figure}

\paragraph*{Global quench from a dimer state. ---} 
 We introduce a simple initial state that breaks time-reversal symmetry  
 \begin{equation}\label{dimer_state}
|  \Psi_0^{(\phi)} \rangle =\bigotimes_{i\, {\rm even}} \frac{ \left( | \uparrow  \downarrow  \rangle_{i ,i+1} + e^{\imath \phi} | \downarrow  \uparrow  \rangle_{i ,i+1}  \right)  }{\sqrt{2}} \epp
 \end{equation}
 This state is not an eigenstate of the XXZ Hamiltonian and is not invariant under spin-flip inversion as it supports a finite spin current 
 on the even sites $\langle \Psi_0^{(\phi)}  | \boldsymbol{J}_{2i} | \Psi_0^{(\phi)}  \rangle = \frac{J}{2} \sin \phi $. The evaluation of the generating functions $\{X_j(\lambda)\}_{j=1}^{\ell-1}$ and $X'(\lambda)$ is straightforward on this state and leads 
 to a complete characterization of its steady state according to (\ref{GGE_even}, \ref{GGE_odd}). 
 Denoting with $v_j(\lambda)$  the effective velocities of the bound states of type $j$ \cite{bonnes14,inho}, 
 the steady state gives in  particular the long-time limit of the spin current 
\begin{equation}
\lim_{t \to \infty} \langle \Psi_0^{(\phi)} (t) | \boldsymbol{J}_i | \Psi_0^{(\phi)} (t)  \rangle = \sum_{j=1}^\ell \int_{-\infty}^{+\infty} d\lambda \rho_j(\lambda) v_j(\lambda) n_j 
\end{equation} 
 which proves to be non-zero and homogeneous on the whole chain 
 due to the presence of the odd conserved charges $\{\boldsymbol{Z}_n\}$ (see Fig.~\ref{fig:time_ev_spin_dimer}). 
 Note how such non-equilibrium protocol probes the non-linear 
 well beyond the regime accessible though the  Drude weight at small $\phi$. 
 { The behavior of the stationary current as a function of $\phi$ is showed in Fig.~\ref{fig:current_dimer}.
 It is interesting to investigate the isotropic limit $\Delta \to 1^-$, where the stationarity of the spin current breaks down: 
 the charges $\{\boldsymbol{Z}_n\}$ become non-local and the steady current vanishes, compatibly with the absence of a 
 finite Drude weight in the isotropic chain \cite{Prosen_vanishing_XXX}. Our data are consistent, for any $\phi$, with a critical form $J(\Delta) \sim (1-\Delta)^\beta$. 

 Finally, we remark that the GGE state obtained from \eqref{GGE_even} without taking into account 
 the charges $\{\boldsymbol{Z}_n\}$,  though correctly predicting expectation values of even local operators, 
 would spoil the odd ones, leading for instance to  a vanishing steady spin current, as pointed out 
 in \cite{Prosen_flux_quench}. 
 
\paragraph*{Joining two states with different spin currents. ---}
We now consider the non-equilibrium steady state \cite{twotemperatures,inho,CAD:hydro, inho2} generated by joining at time $t=0$ 
two dimer states defined in \eqref{dimer_state} with two different phases. Taking $| \Psi_0^{\phi_L} \rangle$ on 
the sites $i \in [-L/2+1,\ldots,0]$ and $ | \Psi_0^{\phi_R} \rangle$ on the sites $i \in [1, \ldots, L/2]$, we consider 
\begin{equation}\label{two_dimers}
|\Psi^{(\phi_L,\phi_R)}_0 \rangle = | \Psi_0^{(\phi_L)} \rangle \otimes  | \Psi_0^{(\phi_R)} \rangle \epc
\end{equation} 
and its time evolution. Using the recent developments on the inhomogeneous quantum quenches \cite{inho,CAD:hydro}, 
combined with our complete characterization of both GGE states, we compute the long-time limit of the local 
magnetization $\langle \boldsymbol{s}^z_i \rangle $ and the local spin current $\langle \boldsymbol{J}_i \rangle $ 
at fixed rays $i= \zeta t$ in the range $-v_L<\zeta<v_R$ with $v_L,v_R$ the maximal velocity of the 
excitations respectively of the left and right steady state \cite{inho,bonnes14}.
Even though the initial state has a uniform zero magnetization, 
we observe the formation of an expanding magnetic domain 
where the magnetization changes 
from $0$ to a finite value which depends on 
the unbalance between the left and right steady-state spin currents
(see Fig. \ref{fig:Profiles_time_ev_spin_dimer}). 
Quite remarkably the same effect appears even when the left and the right state have 
the same initial spin currents, as it is the case when $\phi_R=  \pi - \phi_L$ (see Fig.~\ref{fig:Profiles_time_ev_spin_dimer} bottom). 
This is due to the presence of non-trivial interactions in the model, 
that lead to an asymmetry in the value of the steady state current as function of $\phi$, 
as shown in Fig. \ref{fig:current_dimer}. 
Finally we remark that this intriguing effect is a macroscopic manifestation of the persistent spin 
transport characteristic of the XXZ gapless phase and is a general feature of any junction between 
initial states supporting non-vanishing currents. 
 \paragraph*{Conclusions. ---}
We introduced a complete generalized Gibbs ensemble for the XXZ spin chain that can be extended to any lattice integrable model. We showed how the set of quasi-local charges recently introduced for the gapped regime in \cite{ProsenQLSymmetric} is not sufficient to unambiguously determine the steady state after a quantum quench. 
The existence of additional quasi-local charges had been pointed out in the study of the equilibrium Drude weight \cite{Prosen_Open_QL,PereiraDrude, ProsenDrude}
but their application in the exact non-equilibrium time-evolution had remained up to now elusive. 
We considered the dimer state \eqref{dimer_state} as an explicit example of initial state which breaks time-reversal symmetry 
and displays a steady spin current in the limit $t\to\infty$ which depends non-trivially on the initial phase-shift $\phi$. 
An other interesting example amenable for simple treatment within our framework
would be the quench protocol where a finite magnetic flux is suddenly switched on at $t=0^+$, 
as recently studied on the XXZ spin chain \cite{Prosen_flux_quench,flux_quench} 
as well as in Chern insulators \cite{Chern_Ins}. 
Our analytic approach allowed us to address the XXX limit $\Delta \to 1$ where the steady spin current vanishes
and the behavior close to the isotropic point can be used to estimate the large-time decay.

We also showed how joining two chains generically leads to the spontaneous creation of 
expanding magnetic domains, whose edges are a direct measure of the quasi-particle velocities and which might have
direct connections with the physics of domain growth. 
We remark that once the two states have been realized, only a local quench is necessary to create the junction 
\cite{twotemperatures}~making this protocol amenable for experimental tests.

 \paragraph*{Acknowledgments. ---} %
We are very grateful to Maurizio Fagotti for the collaboration in the early stage of this project.
ADL would like to thank Dario Villamaina for useful discussion on transport in diffusive systems.
This work was supported by by LabEX ENS-ICFP:ANR-10-LABX-0010/ANR-10-IDEX-0001-02 PSL* (J.D.N.), 
the EPSRC Quantum Matter in and out of Equilibrium Ref. EP/N01930X/1 (A.D.L.) and the
European Union's Horizon 2020 research and innovation program 
under the Marie Sklodowska-Curie Grant Agreement No. 701221 (M.C.).


\onecolumngrid
\newpage 

\setcounter{equation}{0}            
\setcounter{section}{0}             
\renewcommand\thesection{\Alph{section}}    
\renewcommand\thesubsection{\arabic{subsection}}    
\renewcommand{\thetable}{S\arabic{table}}
\renewcommand{\theequation}{S\arabic{equation}}
\renewcommand{\thefigure}{S\arabic{figure}}

\begin{center}
{\Large Supplementary Material\\ 
\titleinfo
}
\end{center}
Here we give additional details about the calculations presented in the letter. 
\begin{itemize}
\item In Appendix A we introduce the whole landscape of local and quasi-local conserved quantities of the XXZ model, using all the finite $SU_q(2)$ representations. We recover the integer spin representations, introduced in \cite{ProsenQLSymmetric}, 
and we introduce a family of conserved charges coming from a representation with non-integer spin, 
analogously to what done in \cite{PereiraDrude, ProsenDrude}. 
\item In Appendix B we compute an analytic expression for the eigenvalues of these new charges on the Bethe eigenstates in the thermodynamic limit. 
\item In Appendix C we use the expressions of their eigenvalues to show the relation between the expectation values of these conserved quantities on the initial state and the distribution of rapidities $\{\rho_j\}_{j=1}^\ell$ specifying the GGE steady state. 
\item In Appendix D we show how to evaluate the generating functions $\{ \{X_{2s}\}_{s=\frac{1}{2}}^{ \hat{s}},X'\}$ on the initial state when this is a product spin state. 
\end{itemize}
\section{Family of conserved charges and $SU_q(2)$ representations}
\subsection{Commuting transfer matrices}
We briefly summarize how the XXZ is constructed as an integrable model in the framework
of the algebraic Bethe Ansatz. As mentioned in the text, we rewrite the parameter $\Delta$ of the Hamiltonian as
\begin{equation}
\Delta = \cos \gamma = \frac{q + q^{-1} }{2} \;, \qquad q = e^{\imath \gamma} \;.
\end{equation}
We then introduce the $L$-matrix
defined on the tensor product $V_n \otimes V_a$:
\begin{equation}
\label{LSpin}
 L_{n,a} (\lambda) = \sinh{\lambda} (K_n + K_n^{-1}) \id_a +
 \cosh \lambda (K_n - K_n^{-1}) \sigma_a^z +
 (q-q^{-1}) ( S_n^{-} \sigma_a^{+} + S_n^{+} \sigma_a^{-} )
\end{equation}
where $\lambda$ is the spectral parameter. Here, $V_a = \mathbb{C}^2$ is a associated to a spin $1/2$ representation 
with Pauli spin operators $\sigma_a^{\pm}, \sigma_a^z$, while the space $V_n$ is associated to a representation of the $SU_q(2)$ algebra:
\begin{equation}
\label{SU2qalgebra}
K_n S^{\pm}_n = q^{\pm 1} S^{\pm}_n K_n \;, \qquad [S_n^+, S_n^-] = \frac{K_n^2 - K_n^{-2}}{q - q^{-1}} \;.
\end{equation}
The fundamental representation of dimension $2$ is easily obtained taking 
\begin{equation}
\label{fundrepr}
K_n = q^{\sigma_z/2}\;, \qquad S_n^{\pm} = \frac{1}{2}(\sigma_x \pm \imath \sigma_x) \;.
\end{equation}
In this case the two spaces $V_n \sim V_{a_1} = \mathbb{C}^2$ and $V_a \sim V_{a_2} = \mathbb{C}^2$ have the same dimension
and the matrix $L_{a_1,a_2}(\lambda)$ assumes a symmetric form. One can then introduce the $R$-matrix between them:
\begin{equation} 
R_{a_1,a_2}(\lambda) = L_{a_1, a_2}\bigl(\lambda + \frac{\imath \gamma}{2}\bigr) = 
 \begin{pmatrix}
                                              2 \sinh(\lambda + \imath \gamma) & 0 & 0 & 0\\
                                              0 & 2 \sinh(\lambda) & 2\imath \sin \gamma & 0\\
                                              0 & 2\imath \sin \gamma & 2 \sinh(\lambda) & 0\\
                                              0 & 0 & 0 & 2 \sinh(\lambda + \imath \gamma) 
                                             \end{pmatrix} 
\end{equation}
The matrix $L$ and $R$ satisfy the Yang-Baxter equation in the space $V_{a}\otimes V_{a'} \otimes V_{a_0}$
\begin{equation}
\label{YBstand}
 R_{a, a'}(\lambda - \mu) L_{a,a_0}(\lambda) L_{a,a'_0} (\mu) = L_{a,a'_0} (\mu) L_{a,a_0}(\lambda)  R_{a, a'}(\lambda-\mu) \;.
\end{equation}
In this equation, one interprets the spaces $V_{a}, V_{a'}$ as auxiliary spaces and the space 
$V_{a_0}$ as the quantum space of one single physical spin.
Then, it tells that the $R$-matrix can be used to exchange the $L$-matrices defined on the same physical space.
It is a direct consequence of the algebra in Eq.~\eqref{SU2qalgebra}: in particular generalizations
exist for any pair of auxiliary spaces $V_{n}$ and $V_{n'}$, which allow to exchange $L_{a_0,n}$ and $L_{a_0,n'}$. 

We then introduce the transfer matrix defined as the product of $L$ matrices 
acting on the Hilbert space of a chain of $L$ spin-$1/2$: $V_{a_1}\otimes \ldots \otimes V_{a_L}$
\begin{align}
\label{Tproddef}
 \mathcal{T}_{n}(\lambda) = L_{n, a_1}(\lambda) L_{n, a_2}(\lambda) \cdots L_{n, a_L}(\lambda) = \bigotimes_{i=1}^L L_{n, a_i}(\lambda) \;.
\end{align}
It is easy to verify that Eq.~\eqref{YBstand} (and its generalizations to pairs of representations with $R_{n,n'}(\lambda)$) 
implies an analogous relation for the product of $L$-matrices, the so-called RTT relation
\begin{equation}
\label{RTT}
R_{n, n'}(\lambda - \mu) \mathcal{T}_{n}(\lambda) \mathcal{T}_{n'} (\mu) = \mathcal{T}_{n'} (\mu) \mathcal{T}_{n}(\lambda)  R_{n, n'}(\lambda-\mu) \;. 
\end{equation}
which, upon tracing over the auxiliary spaces, leads to
\begin{equation}
\label{commT}
[T_n(\lambda), T_{n'}(\mu)] = 0 \;, \qquad  T_n(\lambda) = \tr{n}{ \mathcal{T}_n(\lambda)}\;.
\end{equation}
In this way we obtain several families 
of commuting operators in correspondence of each representation of the $SU_q(2)$-algebra in Eq.~\eqref{SU2qalgebra}.

\subsection{Representation of $SU_q(2)$}
Finite dimensional representations of the algebra in \eqref{SU2qalgebra} 
are known and we refer to \cite{sierrabook} for a thorough discussion. 
For generic values of $q$, there is a one-to-one correspondence with the representations of $SU(2)$, 
labeled by the value of the spin $s$. We already showed in Eq.~\eqref{fundrepr} how a two-dimensional representation is triavilly 
obtained from Pauli matrices. In general, the representation of spin $s$ of the undeformed $SU(2)$ has the form
\begin{equation}
 S_s^z \ket{m} = m \ket{m} \;, S_s^{\pm} \ket{m} = \sqrt{(s + 1 \pm m)(s \mp m)} \ket{m \pm 1}\;.
\end{equation}
We can then obtain a corresponding representation of $SU_q(2)$ satisfying the algebra \eqref{SU2qalgebra}
setting
\begin{equation}
\label{SU2qspins}
 K_s = q^{S_s^z}\;, \qquad  S_{s,q}^{\pm} \ket{m} = \sqrt{(s + 1 \pm m)_q(s \mp m)_q} \ket{m \pm 1}
\end{equation}
where the subscript $q$ in $S_{s,q}^{\pm}$ has been added to distinguish them from the undeformed case $S_s^{\pm}$. 
Note that with respect to Eq.~\eqref{SU2qalgebra}, we replaced the abstract subscript $n$ with $s$, indicating the spin value.
Here the notation 
\begin{equation}	
\label{qInt}
 (x)_q = \frac{q^{x} - q^{-x}}{q - q^{-1}}
\end{equation}
has been introduced for the $q$-deformed integers. 

These exhaust the possible irreducible representations of $SU_q(2)$ for generic values of $q$. However,
when $q$ is a root of unity, there are some peculiarity. As explained in the text,
here we focus on the case of principal roots of unity, i.e. $q = e^{\frac{\imath \pi}{\ell}}$. 
One can verify that the representations in \eqref{SU2qspins} are no more irreducible when $s > \smax = \frac{\ell-1}{2}$.
Moreover for the maximal one, i.e. $s = \smax$, there is a class of representations, parameterized by a complex parameter 
$\alpha$, which have the form
\begin{subequations}
\label{PerDefNew}
 \begin{align}
&  K_{\smax, \alpha}\ket{m} \equiv q^{m + \alpha} \ket{m} \;, \qquad m = -\smax,\ldots, \smax \;, \\
&  S_{\smax, \alpha}^{+} \ket{m} = -(m - \smax + 2\alpha)_q \ket{m+1}\;,\\
&  S_{\smax, \alpha}^{-} \ket{m} = (m + \smax)_q \ket{m-1}\;.
 \end{align}
\end{subequations}
Note that for $\alpha \to 0$, this simply reduces to Eq.~\eqref{SU2qspins} for $s = \smax$. 

These representations can be used to build the corresponding transfer matrices 
as explained in \eqref{Tproddef} and \eqref{commT}.
We use the notation $T_{2s}(\lambda) = \tr{2s}{ \mathcal{T}_{2s}(\lambda)}$ for the operator associated with the representation of spin 
$s = 1/2,1,\ldots, \smax = (\ell - 1)/2$. We use instead $T_{2\smax, \alpha}(\lambda)$ 
for the representation defined in \eqref{PerDefNew},
as a function of the parameter $\alpha$. In general we will use $T_{2s}(\lambda)$
to label all of them collectively. 

\section{Eigenvalue of the conserved charges on single-particle eigenstates}
Since all the transfer matrices commute $T_{2s}(\lambda)$ among themselves for any pair of $\lambda,\mu$ and $s,s'$, 
it is possibe to diagonalize all of them simultaneously. 
A special role is played by 
the transfer matrix $\mathcal{T}_1(\lambda)$ associated to the $2$-dimensional fundamental representation. Once written as a $2\times 2$ matrix in the auxiliary space,
it can be expressed in terms of four operators on the quantum space
\begin{equation}
\label{ABCD}
 \mathcal{T}_{1} (\lambda) = \begin{pmatrix}
                              A(\lambda) & B(\lambda) \\ C(\lambda) & D(\lambda)
                             \end{pmatrix} \;.
\end{equation}
so that $T_{1}(\lambda) = A(\lambda) + D(\lambda)$. Then, the relation in Eq.~\eqref{RTT} provides 
the commutation relations between the entries which constitute the Yang-Baxter algebra. 
In particular, one can show that the simultaneous eigenstate of $T_{2s}(\lambda)$ 
can be obtained by the multiple actions of $B(\lambda)$ on the reference state $\ket{\mathbf{0}} = \ket{\uparrow\ldots\uparrow}$
of all spin up
\begin{equation}
 \ket{\lv} = B(\mu_1) \ldots B(\mu_M) \ket{\mathbf{0}}
\end{equation}
provided that the Bethe-Ansatz equations (Eq.~(3) in the main text) are satisfied for the rapidities $\mu_i$.
The eigenvalue of $T_{2s}(\lambda)$ on the state $\ket{\lv}$ will be a symmetric function of the rapidities $\mu_1,\ldots,\mu_M$.
While for $s=1/2$, the eigenvalue can be derived directly derived from the commutation relations deduced 
from Eq.~\eqref{RTT}, for higher spin, the procedure is more involved. 
From the explicit expression of the transfer matrix in Eq.~\eqref{Tproddef} and the
$L$-matrix in Eq.~\eqref{LSpin}, it is easy to obtain the eigenvalue on the reference state
\begin{subequations}
\label{Tvacuum}
\begin{align}
& T_{2s}(\lambda) \ket{\mathbf{0}} = \sum_{m = -s}^s
 (2 \sinh(\lambda + \imath m \gamma))^L\ket{\mathbf{0}} = \sum_{m = -s}^{s} f(\lambda + \imath m \gamma)  \ket{\mathbf{0}} \;, \quad 
 s = \frac{1}{2},\ldots, \smax\\
& T_{2\smax,\alpha}(\lambda) \ket{\mathbf{0}} = \sum_{m = -\smax}^{\smax} f(\lambda + \imath (m - \alpha) \gamma)   \ket{\mathbf{0}} 
 \end{align}
 \end{subequations}
where the function $f(z) = (2\sinh(z))^L$. 
When many $B(\lambda)$ operators act on the reference state In \cite{ProsenQLSymmetric},
this problem was solved for the standard representations by using the fact that higher transfer matrices $T_{2s}(\lambda)$
with $s > 1/2$, can be obtained from the lowest one by tensor product on the auxiliary space. This procedure goes under the name
of ``fusion'' and leads to a functional relation between the transfer matrices with different spin values
\begin{equation}
T_{2s}(\lambda + \frac{\imath \gamma}{2}) 
T_{2s}(\lambda - \frac{\imath \gamma}{2}) =
f\bigl(\lambda + \imath\bigl(s + \frac{1}{2}\bigr) \gamma\bigr)f\bigl(\lambda + \imath\bigl(s + \frac{1}{2}\bigr) \gamma
\bigr) + T_{2s-1}(\lambda) T_{2s+1}(\lambda)\;, \qquad s = \frac{1}{2}, \ldots, \smax  \;.
\end{equation}
An explicit solution for this functional equation can be found as \cite{ProsenQLSymmetric}
\begin{equation}
 \label{TQsol}
 T_{2s}(\lambda) = Q\bigl(\lambda + \imath\bigl(s + \frac{1}{2}\bigr) \gamma\bigr)
 Q\bigl(\lambda - \imath\bigl(s + \frac{1}{2}\bigr) \gamma
\bigr) \sum_{m=-s}^s \frac{f(\lambda + \imath m \gamma )}{Q\bigl(\lambda + \imath\bigl(m + \frac{1}{2}\bigr) \gamma\bigr)
Q\bigl(\lambda + \imath\bigl(m - \frac{1}{2}\bigr) \gamma\bigr)}\;, \qquad s = \frac{1}{2}, \ldots, \smax  \;.
\end{equation}
where $Q(\lambda)$ is the Baxter-$Q$ operator, with eigenvalues:
\begin{equation}
\label{Qeig}
 Q(\lambda) \ket{\lv} = Q(\lambda; \{\mu_1,\ldots, \mu_M \})\ket{\lv} = \prod_{j=1}^M \sinh(\lambda - \mu_j) \ket{\lv} \;. 
\end{equation}
Combining Eqs.~\eqref{TQsol} and \eqref{Qeig}, one obtains the full spectrum $T_{2s}(\lambda, \{\mu_1,\ldots,\mu_M\})$ 
associated to standard representations $s = \frac{1}{2}, \ldots, \smax$ and an arbitrary eigenstate. 
However, a similar approach does not seem to be immediately applicable for the $\alpha$-dependent maximal representation,
as it cannot be obtained by fusing lower-spin ones. 

Here, we follow a different approach. Driven by the simple generalization when $\alpha \neq 0$
in Eqs.~\eqref{Tvacuum}, we assume that $T_{2\smax, \alpha}(\lambda)$ will still be expressed in terms of the $Q$-operator, 
with a structure similar to Eq.~\eqref{TQsol}. In order to find it explicitly, we turn to the simplest possible case:
a state composed by a single rapidity $\mu$. The eigenvalue on this kind of states can be obtained as the ratio
 \begin{equation}
 T_{2s}(\lambda; \{\mu\}) = \frac{\bra{\uparrow\ldots\uparrow \downarrow} T_{2s}(\lambda) B(\mu) \ket{\mathbf{0}}}
 {\bra{\uparrow\ldots\uparrow \downarrow} B(\mu) \ket{\mathbf{0}}} \;.
\end{equation}
This equation is better represented graphically in Fig.~\ref{Fig:transfermatrix}, where the $L$-matrices are represented as boxes
and the contractions over the indexes is indicated as edges joining two boxes.
This representation suggests an efficient way to compute it by performing preliminarly the contractions along 
each physical spin site (e.g. for the numerator the contraction $L_{1/2, 1/2}(\mu)$ with $L_{s,1/2}(\lambda)$ along the
vertical direction); then one is left with a matrix product in the auxiliary space 
(horizontal direction in Fig.~\ref{Fig:transfermatrix}), whose trace can be easily computed by diagonalizing the resulting operator. 
\begin{figure}
 \includegraphics[width=0.7\textwidth]{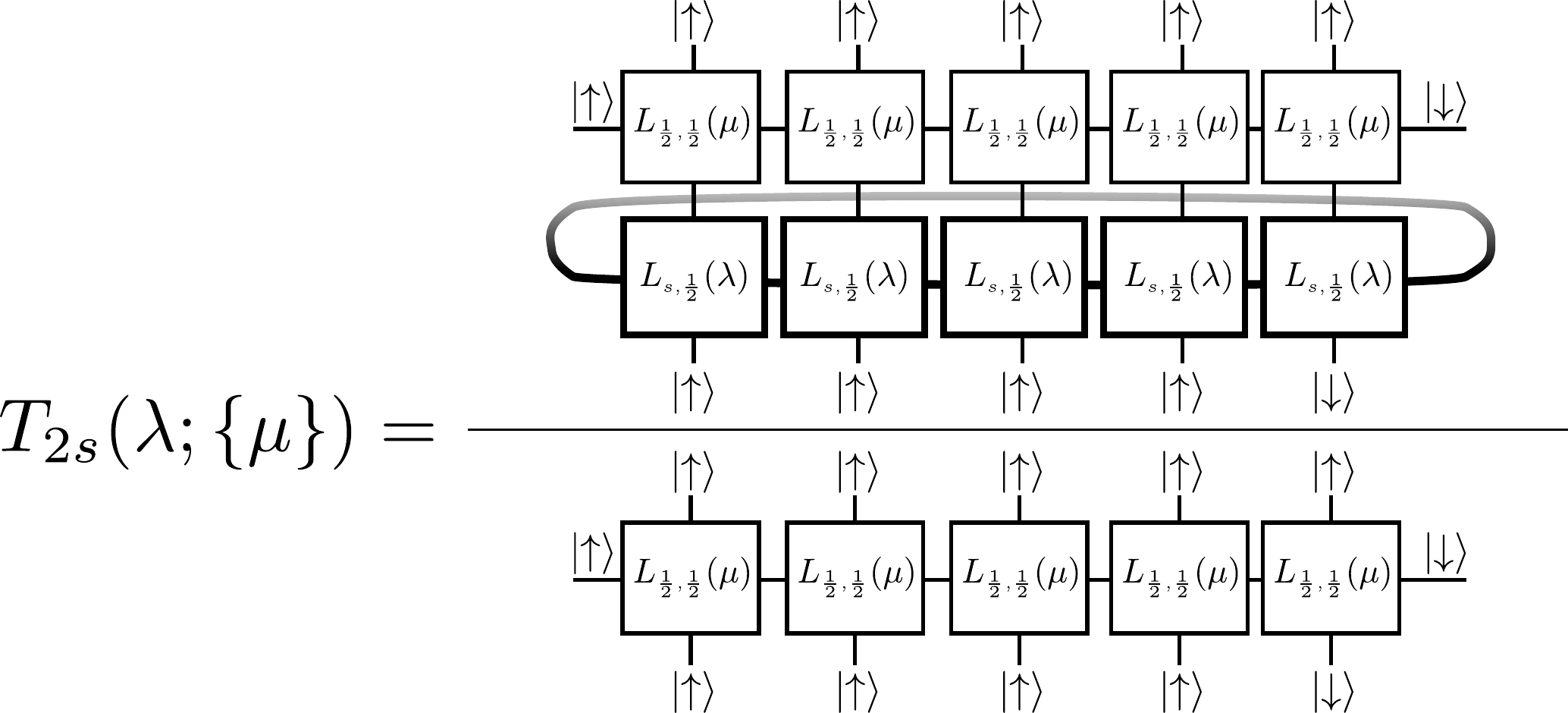}
 \caption{\label{Fig:transfermatrix}
 The eigenvalue of the trace of the transfer matrix over a state with a single rapidity $\mu$ can be written as the ratio of two contractions. 
 The operator $B(\mu)$ can be obtained from the product of $L$-matrices associated with the spin-$1/2$ representation, 
 taking the matrix element between up and down state in the auxiliary space. 
 Similarly, the operator $T_{2s}(\lambda)$ is obtained as product of $L$-matrices associated with the spin-$s$ representation 
 and tracing over the auxiliary space. Reading this equation from left to right, one can compute both the numerator and the denominator
 as product of matrices in the auxiliary spaces. 
 }
\end{figure}
Using that $\mu$ satisfies the Bethe-Ansatz equations (which for a single particle reduce to a quantization condition), 
one obtains finally
\begin{equation}
 \label{TQsolsmax}
 T_{2\smax, \alpha}(\lambda; \{\mu\}) = \sinh\bigl(\lambda + \imath\bigl(\smax - \alpha+ \frac{1}{2}\bigr) \gamma\bigr)
 \sinh\bigl(\lambda - \imath\bigl(\smax - \alpha + \frac{1}{2}\bigr) \gamma
\bigr) \sum_{m=-\smax}^{\smax} \frac{f(\lambda + \imath (m + \alpha) \gamma )}
{\sinh\bigl(\lambda + \imath\bigl(m +\alpha + \frac{1}{2}\bigr) \gamma\bigr)
\sinh\bigl(\lambda + \imath\bigl(m +\alpha - \frac{1}{2}\bigr) \gamma\bigr)}\;.
\end{equation}
As for a single rapidity $\mu$, the eigenvalue of the operator $Q(\lambda)$ are simply given by $\sinh(\lambda - \mu)$,
it is natural to assume that Eq.~\eqref{TQsolsmax} generalizes to an arbitrary number of rapidities promoting $\sinh \to Q$, so
that one arrives at the operator identity:
\begin{equation}
 \label{TQsolsmaxconj}
 T_{2\smax, \alpha}(\lambda) = Q\bigl(\lambda + \imath\bigl(\smax - \alpha+ \frac{1}{2}\bigr) \gamma\bigr)
 Q\bigl(\lambda - \imath\bigl(\smax - \alpha + \frac{1}{2}\bigr) \gamma
\bigr) \sum_{m=-\smax}^{\smax} \frac{f(\lambda + \imath (m + \alpha) \gamma )}
{Q\bigl(\lambda + \imath\bigl(m +\alpha + \frac{1}{2}\bigr) \gamma\bigr)
Q\bigl(\lambda + \imath\bigl(m +\alpha - \frac{1}{2}\bigr) \gamma\bigr)}\;.
\end{equation}
We tested the correctness of this Ansatz numerically for two-rapidities states, but it remains a conjecture
in the general case . 

\section{Relation between conserved quantities and root densities}
The expression \eqref{TQsol} and its generalization \eqref{TQsolsmaxconj}
can be used to obtain the eigenvalues of the trace of the transfer matrix in the thermodynamic limit $L \to \infty$. Indeed,
because of the factor $f(\lambda + \imath m \gamma)$ which involves an $L$-th power, 
for a given value of $\lambda$, only one sector (i.e. a single value of $m$) will exponentially dominate the sum. In particular,
taking $\lambda$ in the neighborhood of the shift-point, i.e.: $\lambda \to  \lambda + \imath \gamma /2$, 
the sum will be dominated by the maximal $m = s$. We therefore define
\begin{align}
\label{Xope}
 &X_{2s}(\lambda) = \frac{1}{2\pi \imath L}\frac{d}{d\lambda} \ln 
 \left[\frac{T_{2s}(\lambda + \imath \gamma/2)} %
 {f(\lambda + \imath \gamma(s + \frac12))}\right] \;, \qquad s = \frac12,\ldots, \smax \\
 & X'(\lambda) = \frac{1}{2\pi \imath L}\left.\frac{d}{d\alpha}\frac{d}{d\lambda} \ln 
 \left[\frac{T_{2\smax,\alpha}(\lambda + \imath \gamma/2)} %
 {f(\lambda+ \imath \gamma(s + \alpha + \frac12))}\right]\right|_{\alpha=0}\;.  \label{Xprimedef}
\end{align}
The operators $X_{2s}(\lambda)$ and $X'(\lambda)$ are generating functions for a complete set of local ($s = 1/2$) and 
quasi-local conserved charges in the XXZ spin chain. In particular, the logarithmic derivative with respect to $\lambda$
ensures that their eigenvalues in each eigenstate will be additive in the rapidities, leading to extensive expectation value
for any operator in the Taylor expansion around $\lambda = 0$ (see Eq.~(10)in the main text).
The factor $f(\ldots)$ in the denominator is chosen so that all of them have a vanishing eigenvalue on the reference state 
$\ket{\mathbf{0}}$, i.e. in absence of any rapidity. 
In the thermodynamic limit, the rapidities in any eigenstate are arranged according to the string hypothesis~\cite{Takahashibook}:
\begin{align}
\lambda_{p, a}^{n,\upsilon} = \lambda_{p}^{n,\upsilon} 
+ \frac{\imath \gamma}{2} (n + 1 - 2a) + \frac{\imath \pi(1 - \upsilon)}{4} 
+ \imath \delta_{p}^{n,a} \;, a = 1,\ldots, n \;.
\end{align}
In this expression, the index $a$ labels the rapidities 
belonging to the same string, with the same real part $\lambda_{p}^n$; 
the $\delta_{p}^{n,a}$ 
is the deviation from the string hypothesis, which becomes exponentially small in the system size $L$. String types 
are identified by the pair $(n, v)$, with $n$ the size of the string (number of rapidities)
and $\upsilon = \pm 1$ is the string parity. In the gapless regime, whenever $\Delta = \cos \gamma$, with $\gamma$ 
is a rational multiple of $\pi$ there exist only a finite number of string types. In particular,
for the simplest case considered here, $\gamma = \pi/\ell$, one has exactly $\ell$ different types, $j=1,\ldots,\ell$
with $n_j$ and $\upsilon_j$   
\begin{equation}
n_j = \begin{cases} 
	j\;, & j = 1,\ldots, \ell-1\\
        1\;, & j = \ell
       \end{cases} \;, \qquad
\upsilon_j = \begin{cases} 
	\;1\;, & j = 1,\ldots, \ell-1\\
        -1\;, & j = \ell
       \end{cases}
\end{equation}

In the thermodynamic limit the string momenta $\lambda_\alpha^{(j)}$ become dense on the real line and for each string type $j$ we can introduce a density distribution $L \rho_j(\lambda_\alpha^{(j)}) = (\lambda_{\alpha+1}^{(j)}-\lambda_\alpha^{(j)})^{-1}$ of them as a set of occupied (particles) and unoccupied (holes) root distributions $\{\rho_j\}_{j=1}^{N_s} \cup \{ \rho^h_{j}\}_{j=1}^{N_s}$, one for each string type. 
The two set of distributions are related by the thermodynamic version of the Bethe equations reading as
\begin{equation} \label{Eq:TBA}
\upsilon_j \rho^t_j(\lambda) = { a}_j(\lambda)- \sum_{k=1}^{N_s}\int{\rm d}\mu\, T_{jk}(\lambda-\mu)\rho_{k}(\mu)  \quad\quad j = 1,\ldots, N_s \,,
\end{equation}
where $\rho^t_j(\lambda)\equiv\rho_j(\lambda) + \rho^h_{j}(\lambda)$. Therefore only one of two sets, either the density of particles or of the holes, is sufficient to completely characterize the state in the thermodynamic limit.  
The sum over $k$ runs over all the possible $N_s$ types of particles with different parities $\{ \upsilon_j\}_{j=1}^{N_s}$ and lengths $\{ n_j\}_{j=1}^{N_s}$. 
We can then write the scattering kernels in \eqref{Eq:TBA} as
\begin{align}\label{kernels}
&a_{n,\upsilon}(\lambda) =
 \frac{\upsilon}{\pi} \frac{\sin(\gamma n)}{\cosh(2\lambda) - \upsilon \cos(\gamma n)} \epc  \\
 &T_{jk}(\lambda) =
(1-\delta_{n_j n_k}) a_{|n_j-n_k|,\upsilon_j \upsilon_k}(\lambda)   
+ 2a_{|n_j-n_k|+2,\upsilon_j \upsilon_k}(\lambda) 
&+ ... + 
2a_{n_j+n_k-2,\upsilon_j \upsilon_k} (\lambda) + a_{n_j+n_k,\upsilon_j \upsilon_k} (\lambda)\,. 
\end{align}
where $a_j  (\lambda) \equiv a_{n_j,\upsilon_j}  (\lambda)$.  \\

The conserved quantities can be evaluated on a thermodynamic state by considering the thermodynamic limit of their eigenvalue. 
These read as 
\begin{subequations}
\label{Xrho}
\begin{align}
& X_{2s}(x) \ket{\rho} = \sum_{j=1}^{\ell} \int_{-\infty}^\infty d \lambda \; 
q^{(s)}_{j}(x - \lambda) \rho_j(\lambda) \ket{\rho} \;, \qquad s = \frac12,1,\ldots,\smax \label{Xeven}\\
& X'(x) \ket{\rho} = \sum_{j=1}^{\ell} \int_{-\infty}^\infty d \lambda \; q_{j}'(x - \lambda) \rho_j(\lambda) \ket{\rho}  \label{Xprime}
\end{align}
\end{subequations}
In these expressions, the eigenvalues $q^{(s)}_{j}(x)$ (and $q_{j}'(x-\lambda)$) 
corresponding to each $(n_j, \upsilon_j)$-string
are obtained from those of a single rapidity, by summing over the whole string:
\begin{align}
& q^{(s)}_{1} (x) = \frac{1}{2\pi \imath}\frac{d}{dx} \ln \left[
\frac{\sinh(x - \frac{\imath s \pi}{l})}{\sinh(x + \frac{\imath s \pi}{l})}
\right] = a_{2s,1}(x) \;, \quad s = \frac12,1,\ldots, \smax \;,\\
& q^{(s)}_{j}(x) =  \sum_{a=1}^{n_j} q^{(s)}_{1}\Bigl(x + \frac{\imath \gamma (n_j + 1 - 2a)}{2} + \frac{\imath \pi (\upsilon_j -1)}{4}\Bigr) = \sum_{a=1}^{\min\{n_j, 2s\}}
a_{|2s - n_j| -1 + 2a, \upsilon_j}(x) \;. \label{q2sj}
\end{align} 
In a similar way, we can treat $q'_j(x)$ and obtain
\begin{align}
q' (x) = 
\left.\frac{d}{d\alpha}a_{2(\smax - \alpha), 1}(x) \right|_{\alpha=0}
\;,
\qquad 
q_j'(x) = \left.\frac{d}{d\alpha}\sum_{a=1}^{n_j} a_{2 (\smax-\alpha) - n_j - 1 + 2a,\upsilon_j}(x)\right|_{\alpha = 0}
\end{align}
where we introduced the function
\begin{equation}
a_{n,\upsilon}(\lambda) = \frac{\upsilon}{\pi} \frac{\sin(\gamma n)}{\cosh(2\lambda) - \upsilon \cos(\gamma n)} 
\end{equation}

Eqs.~\eqref{Xrho} relates the expectation value of the full set of conserved quantities with the distributions of rapidities
$\{\rho_1,\ldots,\rho_\ell\}$. The $\ell-1$ generating functions of even charges $X_{2s}(x)$, with $s = \frac12,1,\ldots,\smax$
can be used to fix $\rho_1,\ldots, \rho_{\ell-2}$ and the difference $\rho_{\ell} - \rho_{\ell-1}$.  
This is a consequence of the following relations between the eigenvalues
\begin{equation}
 q^{(s)}_{\ell}(x) = - q^{(s)}_{\ell-1}(x) \;, \qquad s = \frac12,1,\ldots,\smax
\end{equation}
which can explicitly checked in Eq.~\eqref{q2sj}.
To simplify the notation we rearrange the $\ell$ functions $\rho_j$, $j = 1,\ldots, \ell$ as
\begin{equation}
 \tilde{\rho}_j(\lambda) = \begin{cases}
                            \rho_j(\lambda) & j = 1,\ldots, \ell-2\\
                            \rho_{\ell-1}(\lambda) - \rho_{\ell}(\lambda) & j = \ell-1\\
                            \rho_{\ell}(\lambda) & j = \ell
                           \end{cases}
\end{equation}
and we can rewrite Eq.~\eqref{Xeven} as
\begin{equation}
\label{Xeven1}
X_{2s}(x) \ket{\rho} = \sum_{j=1}^{\ell-1} \int_{-\infty}^\infty d \lambda \; q^{(s)}_{ j}(x - \lambda) \tilde{\rho}_j(\lambda) \ket{\rho} \;, \qquad s = \frac12,1,\ldots,\smax \;.
\end{equation}
We can now invert these relations.
In order to do so, we observe that Eqs.~\eqref{Xrho} and \eqref{Xeven1} have 
the form of convolutions and becomes therefore multiplicative when going in Fourier transform. 
Defining for any function $g(x)$, the Fourier transform $\hat g(p)$ as
\begin{equation}
 \hat g(p) = \int_{-\infty}^\infty dx \, g(x) e^{\imath p x}
\end{equation}
we have for the function $a_{n,v}(\lambda)$ the expression
\begin{equation}
\label{atransf}
 \hat a_{n,v}(p) =  \frac{\sinh \left(\frac{\pi k_v p}{2} -\frac{p \gamma  n}{2}\right)}{\sinh \left(\frac{\pi  p}{2}\right)} 
 \;, \qquad  k_{v} = \begin{cases}
          2\lfloor \frac{n \gamma}{2 \pi} \rfloor + 1 & v = 1\\
          2\lfloor \frac{n \gamma}{2 \pi} + \frac 12 \rfloor & v = -1
         \end{cases}
 \end{equation}
which holds for any $n \in \mathbb{R}$ and $v = \pm 1$.
Taking the Fourier transform of $q^{(s)}_{j}(x)$ in Eq.~\eqref{q2sj}, we have
\begin{equation}
\label{resumcosh}
\hat{q}^{(s)}_{j}(p) \equiv \frac{\cosh\bigl(\frac{\ell p \pi - p \pi |j - 2 s|}{2 \ell}\bigr)-
\cosh\bigl(\frac{\ell p \pi - p \pi (2s + j)}{2 \ell}\bigr)}{2 \sinh(\frac{p \pi}{2 \ell}) \sinh(\frac{p \pi}{2})}
\end{equation}
Instead, in order to invert the relation \eqref{Xprime} involving $X'(x)$, we need the Fourier transform of $q_j'(x)$
\begin{equation}
q_j'(p) = \left.\frac{d}{d\alpha}\sum_{a=1}^{n_j} 
  \hat{a}_{\ell - n_j + 2(a-1) + 2\alpha,\upsilon_1}(p)\right|_{\alpha=0}= 
\begin{cases}
- \frac{\pi p \sinh \bigl(\frac{\pi  j p}{2\ell}\bigr)}{\ell \sinh(\frac{\pi p}{2}) \tanh \bigl(\frac{\pi  p}{2\ell}\bigr) }  & j = 1,\ldots, \ell-1\\
- \frac{\pi p \cosh\bigl(\frac{\pi (\ell - 1)p}{2\ell}\bigr)}{\ell \sinh(\frac{\pi p}{2})} & j = \ell
\end{cases}
\end{equation}
Finally, using the recursion relation
\begin{equation}
 \hat{q}^{(s+1/2)}_{j}(p) + 
 \hat{q}^{(s-1/2)}_{j}(p) - 2 \cosh\Bigl(\frac{p \pi}{2 l}\Bigr) \hat{q}^{(s)}_{j}(p) = - \sinh\Bigl(\frac{\pi p}{2}\Bigr)
 \delta_{j,2s}
\end{equation}
we arrive at the final set of equations for the $\hat{\rho}_{j}(p)$:
\begin{align}
& \label{rhosol}
 \hat{\rho}_{j}(p) - \delta_{j, \ell-1} \, \hat{\rho}_{\ell}(p) = 2 \cosh\bigl(\frac{\pi p}{2 \ell}\bigr) \hat{X}_{j}(p) -  \hat{X}_{j+1}(p) 
 -   \hat{X}_{j-1}(p)  \;, \quad j = 1,\ldots, \ell - 1\\
& \label{finalrhoell}
 \hat{\rho}_{\ell}(p) = - \cosh\bigl(\frac{\pi p}{2 \ell}\bigr) \hat{X}_{\ell -1}(p) 
 - \frac{\ell}{\pi p} \sinh\bigl(\frac{\pi p}{2 \ell}\bigr)\hat{X}'(p) 
\end{align}

These equations allow to fix a representative eigenstate $\ket{\rho}$
in terms of the expectation values of all the local and quasi-local charges, generated by $X_{2s}(x)$ and $X'(x)$.
In this way, as $X_{2s}(x)$ and $X'(x)$ remain constant througout the quantum dynamics, they can be computed on the initial
state and used to obtain the microcanonical GGE described by the corresponding $\ket{\rho}$. This gives the complete characterization of any GGE state in the XXZ chain. 

\subsection{Spin-flip invariant GGE}
Here we show that if the if the initial state $\ket{\Phi}$ has a definite parity under spin-flip,
i.e. 
\begin{equation}
\label{Sapplication}
\mathcal{S} \ket{\Phi} = \pm \ket{\Phi} \;.
\end{equation}
where the spin-flip operator $\mathcal{S} = \left(\prod_{i=-\frac{L}{2}}^{\frac{L}{2}} \boldsymbol{s}^x_i \right) $ 
as already introduced in the main text, then the constraint \eqref{finalrhoell} reduces to $\rho_{\ell} = \rho_{\ell-1}^h$ . Under the spin-flip, the transfer matrix satisfies
\begin{equation}
\mathcal{S} T_{2\smax, \alpha} (\lambda) \mathcal{S} = T_{2\smax, -\alpha} (\lambda)  \;.
\end{equation}
From the definition of $X'(x)$ in \eqref{Xprimedef}, we then deduce that
\begin{equation}
X'(x) + \mathcal{S} X'(x) \mathcal{S} = 
\frac{1}{2\pi \imath L}\frac{d}{dx} \ln \left[ 
\frac{f(x + \imath \gamma(s - \alpha + \frac12))}{f(x + \imath \gamma(s + \alpha + \frac12))} 
 \right] 
\end{equation}
Then using Eq.~\eqref{Sapplication}, 
we have for the expectation value of \eqref{Xprimedef} on $\ket{\Phi}$ 
\begin{equation}
\label{XpSpinInv}
X'_s(x; \Phi) \equiv \bra{\Phi} X'_s \ket{\Phi} = - \frac{1}{2\ell (\cosh x)^2}\;, \qquad \ket{\Phi}\text{ eigenstate of }\mathcal{S}
\end{equation}
Now we recall that from the BA equations (equation (3) in the main text) one has
\begin{equation}
 \rho_j^t(x) = a_j(x) - X_{j+1}(x) - X_{j-1}(x) \;, \qquad j = 1,\ldots, \ell-1
\end{equation}
and taking the difference with \eqref{rhosol} leads simply to 
\begin{equation}
 \rho^h_j(x) + \delta_{j, \ell-1} \rho_{\ell}(x) = a_j(x) - 
 X_{j}(x + \frac{\imath \gamma}{2}) - X_{j}(x - \frac{\imath \gamma}{2}) \;, \qquad j = 1,\ldots,\ell-1  \;,\\
\end{equation}
Note here the additional term for $j= \ell-1$, which simply comes from $\rho'_{\ell-1} = \rho_{\ell-1} - \rho_{\ell}$.
For $j = \ell-1$, going to Fourier transform we have
\begin{equation}
\label{rhohrhol}
 \hat\rho^h_{\ell-1}(p) + \hat{\rho}_{\ell}(p)= \hat a_{\ell-1}(p) - 2\cosh(\frac{\pi p}{2 \ell}) \hat{X}_{j}(p)
\end{equation}
Now, using \eqref{XpSpinInv}, 
and \eqref{atransf} (we recall that $a_j = a_{n_j, \upsilon_j}$ and $n_{\ell-1} = \ell-1$ and $\upsilon_{\ell-1} = 1$), 
we see that for any $\ket{\Phi}$ with definite spinflip parity, we have:
\begin{equation}
\label{aelleven}
 \hat a_{\ell -1} (p) = \frac{\sinh\Bigl(\frac{\pi p}{2 \ell}\Bigr)}{\sinh\Bigl(\frac{\pi p}{2}\Bigr)}=
 - \frac{2\ell}{\pi p} \sinh\Bigl(\frac{\pi p}{2\ell}\Bigr) \hat{X}'(p; \Phi) \;.
\end{equation}
Finally, inserting \eqref{aelleven} in \eqref{rhohrhol} and comparing the resulting expression with \eqref{finalrhoell}, we see that for any
state with definite parity we have
\begin{equation}
 \rho_{\ell-1}^h + \rho_{\ell} = 2 \rho_{\ell} \quad \Rightarrow \quad  \rho_{\ell} = \rho_{\ell-1}^h 
\end{equation}

\section{Evaluation of the conserved charges on a generic product state}
It is easy to show that the decomposition of the transfer matrix $T_{2s}$ into a product of $L$ operators as in \eqref{Tproddef} allows to compute the generating functions \eqref{Xprimedef} on a generic product state
\begin{equation}
| \Phi \rangle = \bigotimes_{i} | \Phi_{i,i+p} \rangle
\end{equation}
where $| \Phi_{i,i+p} \rangle$ is a spin state for the spins between the position $i$ and $i+p$ with $p$ generic integer. Following \cite{IlievskiJSTAT} we introduce the $(2s +1) \times (2s + 1)$ matrix of operators acting on the the space $\mathbb{C}^2$ of the spin in position $i$  
\begin{equation}
\mathbb{L}_{2s,i}(x,\mu)= \frac{L_{2s,a_i} (x) L_{2s,a_i}^*(x+\mu)}{\sinh(x + \imath \gamma ( s + \frac{1}{2}) )  \sinh(x+\mu - \imath \gamma ( s + \frac{1}{2}) ) } \;.
\label{eqn:two_channel_Lax}
\end{equation}
where the product of the two Lax operators (defined in \eqref{LSpin}) is taken with respect to the auxiliary space indices. With this we define the following expectation value on one single constituent $| \Phi_{i,i+p} \rangle$ of the product state 
\begin{equation}
\mathbb{T}^{\Phi}_{2s}(x,\mu) =  \langle  \Phi_{i,i+p} |  \mathbb{L}_{2s,i}(x,\mu)\cdots \mathbb{L}_{2s,i+p}(x,\mu)| \Phi_{i,i+p} \rangle,
\end{equation}
which is still a matrix for dimension $(2s +1) \times (2 s + 1)$. Finally we define the generating function as traces over the auxiliary space indices 
\begin{equation}
X_{2s}(x) =\frac{1}{2\pi \imath p}
\frac{{\rm Tr}\left({\rm Adj}(\mathbb{T}^{\Phi}_{2s}(x,0)-1)\mathbb{D}^{\Phi}_{2s}(x,0)\right)}
{{\rm Tr}\left({\rm Adj}(\mathbb{T}^{\Phi}_{2s}(x,0)-1)\right)},
\end{equation}
where $\mathbb{D}^{\Phi}_{2s}(\mu,x)=\partial_{\mu}\mathbb{T}^{\Phi}_{2s}(x,\mu)|_{\mu=0}$ and
the matrix coadjoint is defined as ${\rm Adj}(A)\equiv \det{(A)}A^{-1}$. The same can be done for the generating function $X'(x)$ simply by introducing the also the derivative respect to $\alpha$. 
We define
\begin{equation}
\mathbb{L}_{2\hat{s}, \alpha,i}(x,\mu)= \frac{L_{(2\hat{s},\alpha),a_i} (x) L_{(2\hat{s},\alpha),a_i}^*(x+\mu)}{\sinh(x + \imath \gamma ( \hat{s} + \alpha+ \frac{1}{2}) )  \sinh(x+\mu - \imath \gamma ( \hat{s} - \alpha + \frac{1}{2}) )} .
\label{eqn:two_channel_Lax}
\end{equation}
such that 
\begin{equation}
\mathbb{T}^{\Phi}_{2\hat{s},\alpha}(x,\mu) =  \langle  \Phi_{i,i+p} |  \mathbb{L}_{2\hat{s},\alpha,i}(x,\mu)\cdots \mathbb{L}_{2\hat{s},\alpha,i+p}(x,\mu)| \Phi_{i,i+p} \rangle,
\end{equation}
With these elements we can then define the generating function $X'(x)$
\begin{equation}
X' (x) =\frac{1}{2 \pi \imath p}\frac{d}{d\alpha}\left[
\frac{{\rm Tr}\left({\rm Adj}(\mathbb{T}^{\Phi}_{2\hat{s},\alpha}(x,0)-1)\mathbb{D}^{\Phi}_{2\hat{s},\alpha}(x,0)\right)}
{{\rm Tr}\left({\rm Adj}(\mathbb{T}^{\Phi}_{2\hat{s},\alpha}(x,0)-1)\right)}\right]_{\alpha = 0},
\end{equation}
with $\mathbb{D}^{\Phi}_{2s,\alpha}(\mu,x)=\partial_{\mu}\mathbb{T}^{\Phi}_{2\hat{s},\alpha}(x,\mu)|_{\mu=0}$.\\

Note that  the same method can be applied to evaluate the generating functions $X_{2s}$ and $X'$ when $| \Phi \rangle$ is a more generic matrix product state. 

%
%


\begin{thebibliography}{99}
%

\bibitem{expXXZ}
A. V. Sologubenko, T. Lorenz, H. R. Ott and J. Low Temp. Phys. \href{\doi10.1007/s10909-007-9317-x}{\bf 147} 387–403 (2007); 
O. Breunig \emph{et al}, Phys. Rev. Lett. \href{http://link.aps.org/doi/10.1103/PhysRevLett.111.187202}{\bf 111}, 187202 (2013);  A. Freimuth , L. S. Wu \emph{et al}, Science \href{\doi10.1126/science.aaf0981}{\bf 352} , 1206-1210 (2016); M. Mourigal \emph{et al}, Nature Physics \href{\doi10.1038/nphys2652}{ \bf 9}, 435-441 (2013); J. Schlappa \emph{et al}, Nature \href{\doi10.1038/nature10974}{\bf 485}, 82-85 (2012); B. Lake, D. A. Tennant, J.-S. Caux, T. Barthel, U. Schollw\"ock, S. E. Nagler, and C. D. Frost, Phys. Rev. Lett. \href{10.1103/PhysRevLett.111.137205}{\bf 111}, 137205 (2013).

\bibitem{exp_unitary_time_evolution}
M. Greiner \emph{et al}, 
Nature \href{\doi10.1038/nature00968}{\bf 419}, 51-54 (2002); T. Kinoshita, T. Wenger,  and D. S. Weiss, 
 Nature \href{\doi10.1038/nature04693}{\bf 440}, 900 (2006);
 S. Hofferberth, I. Lesanovsky \emph{et al}, 
Nature \href{\doi10.1038/nature06149}{\bf 449}, 324-327 (2007); 
L. Hackermuller, U. Schneider \emph{et al}, 
Science \href{\doi10.1126/science.1184565}{\bf 327}, 1621 (2010);

\bibitem{exp_transport}
T. Fukuhara, A. Kantian \emph{et al}, 
Nature Physics \href{\doi10.1038/nphys2561}{\bf 9}, 235 (2013); T. Fukuhara, P. Schau{\ss} \emph{et al}, 
Nature \href{\doi10.1038/nature12541}{\bf 502}, 76 (2013);
J.P. Ronzheimer, M. Schreiber \emph{et al}, 
Phys. Rev. Lett. \href{http://dx.doi.org/10.1103/PhysRevLett.110.205301}{\bf 110}, 205301 (2013);
U. Schneider, L. Hackerm\"uller \emph{et al}, 
Nature Phys. \href{\doi10.1038/nphys2205}{\bf 8}, 213 (2012); 
M. Cheneau, P. Barmettler \emph{et al}, 
Nature \href{http://dx.doi.org/10.1038/nature10748}{\bf 481}, 484 (2012); 
P. Jurcevic, B. P. Lanyon \emph{et al}, 
Nature \href{http://dx.doi.org/10.1038/nature13461}{\bf 511}, 202 (2014).


\bibitem{exp} 
S. Trotzky, Y.-A. Chen \emph{et al}, 
Nature Phys. \href{\doi10.1038/nphys2232}{\bf 8}, 325 (2012); M. Gring, M. Kuhnert \emph{et al}, 
Science \href{\doi10.1126/science.1224953}{\bf 337}, 1318 (2012):
T. Langen, R. Geiger \emph{et al}, 
Nature Physics \href{http://dx.doi.org/10.1038/nphys2739}{\bf 9}, 640 (2013);
F. Meinert, M.J. Mark \emph{et al}, 
Phys. Rev. Lett. \href{http://dx.doi.org/10.1103/PhysRevLett.111.053003}{\bf 111}, 053003 (2013); 

\bibitem{EF:review} F.H.L. Essler and M. Fagotti, J. Stat. Mech. (2016) \href{\doi10.1088/1742-5468/2016/06/064002}{064002}; P. Calabrese and J. Cardy J. Stat. Mech. (2016) \href{http://iopscience.iop.org/article/10.1088/1742-5468/2016/06/064003}{064003}; J.-S. Caux J. Stat. Mech. (2016) \href{http://iopscience.iop.org/article/10.1088/1742-5468/2016/06/064006}{064006}; L. Vidmar and M. Rigol J. Stat. Mech. (2016) \href{http://iopscience.iop.org/article/10.1088/1742-5468/2016/06/064007}{064007l}; L. D'Alessio, Y. Kafri, A. Polkovnikov, and M. Rigol, Adv. Phys. \href{http://www.tandfonline.com/doi/full/10.1080/00018732.2016.1198134}{\bf 65}, 239 (2016); C. Gogolin and J. Eisert, Rep. Prog. Phys. {\bf 79}, \href{\doi10.1088/0034-4885/79/5/056001}{056001} (2016); A. De Luca and G. Mussardo, J. Stat. Mech. (2016) \href{\doi10.1088/1742-5468/2016/06/064011}{064011}; 

 




\bibitem{dephasing} T. Barthel and U. Schollw\"ock, 
Phys. Rev. Lett. \href{http://dx.doi.org/10.1103/PhysRevLett.100.100601}{\bf 100}, 100601 (2008).
%
%

\bibitem{quench-xxz-fagotti}
M. Fagotti and F. H. L. Essler, 
J. Stat. Mech. (2013) \href{http://dx.doi.org/10.1088/1742-5468/2013/07/P07012}{P07012}; . Fagotti, M. Collura,  F. H. L. Essler, and P. Calabrese,  
Phys. Rev. B \href{http://dx.doi.org/10.1103/PhysRevB.89.125101} {\bf 89}, 125101 (2014); 

\bibitem{quench-xxz}
B. Pozsgay, J. Stat. Mech. (2013) \href{http://dx.doi.org/10.1088/1742-5468/2013/07/P07003}{P07003};  M
B. Wouters, J. De Nardis \emph{et al}, 
Phys. Rev. Lett.  \href{http://dx.doi.org/10.1103/PhysRevLett.113.117202}{\bf 113}, 117202 (2014);
B. Pozsgay, M. Mesty\'{a}n \emph{et al}, 
Phys. Rev. Lett. \href{http://dx.doi.org/10.1103/PhysRevLett.113.117203}{{\bf 113}}, 117203 (2014); L. Piroli, B. Pozsgay, E. Vernier, \href{https://arxiv.org/abs/1611.06126}{arXiv:1611.06126} (2016);
 L. Piroli, E. Vernier, P. Calabrese and M. Rigol, \href{https://arxiv.org/abs/1611.08859}{arXiv:1611.08859} (2016); 
  E. Ilievski, E. Quinn, J-S Caux, \href{https://arxiv.org/abs/1610.06911}{arXiv:1610.06911} (2016).





\bibitem{Piroli_XXZ}
L. Piroli, E. Vernier, and P. Calabrese, Phys. Rev. B \href{10.1103/PhysRevB.94.054313}{\bf 94}, 054313 (2016).



\bibitem{IlievskiPRLQL}
E. Ilievski, J. De Nardis, B. Wouters, J-S Caux, F. H. L. Essler, T. Prosen
Phys. Rev. Lett. \href{\doi10.1103/PhysRevLett.115.157201}{\bf 115}, 157201 (2015).
\bibitem{IlievskiJSTAT}
 E. Ilievski, E. Quinn, J. De Nardis, M. Brockmann J. Stat. Mech. (2016) \href{\doi10.1088/1742-5468/2016/06/063101}{063101}.

\bibitem{GGEcit}
M. Rigol, V. Dunjko and M. Olshanii, Nature  \href{http://www.nature.com/nature/journal/v452/n7189/full/nature06838.html}{\bf 452}, 854-858 (2008); 
M. Rigol, V. Dunjko, V. Yurovsky, and M. Olshanii, Phys. Rev. Lett. \href{http://link.aps.org/doi/10.1103/PhysRevLett.98.050405}{\bf 98}, 050405 (2007); 
F. H. L. Essler, S. Evangelisti, and M. Fagotti, 
Phys. Rev. Lett.  \href{http://dx.doi.org/10.1103/PhysRevLett.109.247206}{\bf 109}, 247206 (2012).


\bibitem{Prosen_Open}
T. Prosen, Phys. Rev. Lett. \href{\doi10.1103/PhysRevLett.106.217206}{\bf 106}, 217206 (2011),
T. Prosen, Phys. Rev. Lett. \href{\doi10.1103/PhysRevLett.107.137201}{\bf 107}, 137201 (2011),
M. Znidaric, B. Zunkovic, T. Prosen,  Phys. Rev. E \href{\doi10.1103/PhysRevE.84.051115}{\bf 84}, 051115 (2011).








\bibitem{inho}
B. Bertini, M. Collura, J. De Nardis, M. Fagotti, 	Phys. Rev. Lett. \href{\doi10.1103/PhysRevLett.117.207201}{\bf 117}, 207201 (2016).

\bibitem{inho2}
L. Vidmar, D. Iyer, M. Rigol, 	\href{https://arxiv.org/abs/1512.05373}{arXiv:1512.05373} (2016); 
W. Xu, M. Rigol 	\href{https://arxiv.org/abs/1612.08988}{arXiv:1612.08988} (2016)

\bibitem{Hirobe}
D. Hirobe \textit{et al}, Nature Physics \href{http://www.nature.com/nphys/journal/v13/n1/full/nphys3895.html}{\bf 13}, 30--€"34 (2017)

\bibitem{Prosen_Open_QL}
T. Prosen, E. Ilievski, Phys. Rev. Lett. \href{\doi10.1103/PhysRevLett.111.057203}{\bf 111}, 057203 (2013).




\bibitem{ProsenDrude}
T. Prosen,  Nuclear Physics, Section B \href{\doi10.1016/j.nuclphysb.2014.07.024}{\bf 886}, 1177 (2014);
T. Prosen, Nucl. Phys. B \textbf{886}, 1177 (2014).

\bibitem{PereiraDrude}
R. G. Pereira, V. Pasquier, J. Sirker, I. Affleck  J. Stat. Mech. (2014) P09037.







\bibitem{Drude_XXZ_preProsen}
X. Zotos, Phys. Rev. Lett. \href{\doi/10.1103/PhysRevLett.82.1764}{\bf 82}, 1764 (1999);
J. Herbrych, P. Prelovšek, and X. Zotos, Phys. Rev. B \href{10.1103/PhysRevB.84.155125}{\bf 84}, 155125 (2011);
M. \v{Z}nidari\v{c}, A. Scardicchio, and V. K. Varma, Phys. Rev. Lett. \href{\doi/10.1103/PhysRevLett.117.040601}{\bf 117}, 040601 ;
G. Majumder and A. Garg, Phys. Rev. B \href{10.1103/PhysRevB.94.134508}{\bf 95}, 134508;
J. M. P. Carmelo, T. Prosen, and D. K. Campbell, Phys. Rev. B \href{\doi/10.1103/PhysRevB.92.165133}{\bf 92}, 165133;
R. Steinigeweg, J. Gemmer, and W. Brenig, Phys. Rev. B \href{\doi/10.1103/PhysRevB.91.104404}{\bf 91}, 104404 ;
C. Karrasch, J. Hauschild, S. Langer, and F. Heidrich-Meisner, Phys. Rev. B \href{\doi/10.1103/PhysRevB.87.245128}{\bf 87}, 245128; 
C. Karrasch, J. H. Bardarson, and J. E. Moore, Phys. Rev. Lett. \href{\doi/10.1103/PhysRevLett.108.227206}{\bf 108}, 227206; J. Herbrych, R. Steinigeweg, and P. Prelovšek, Phys. Rev. B \href{10.1103/PhysRevB.86.115106}{\bf 86}, 115106;
J. Sirker, R. G. Pereira, and I. Affleck, Phys. Rev. Lett. \href{\doi/10.1103/PhysRevLett.103.216602}{\bf 103}, 216602 (2009); C. Karrasch, \href{https://arxiv.org/abs/1611.00573}{arXiv:1611.00573} (2016). 

\bibitem{Prosen_Prosal}
C. Karrasch, T. Prosen and F. Heidrich-Meisner, 	\href{https://arxiv.org/abs/1611.04832}{arXiv:1611.04832} (2016).


\bibitem{Takahashibook} 
M. Takahashi, {\em {Thermodynamics of One-dimensional Solvable Models}}
 (Cambridge University Press, 2005).


\bibitem{Korepinbook}
V.E. Korepin, A.G. Izergin, and N.M. Bogoliubov, {\em {Quantum Inverse
  Scattering Method, Correlation Functions and Algebraic Bethe Ansatz}}
  (Cambridge University Press, 1993).



\bibitem{ETH}
J. M. Deutsch, Phys. Rev. A \href{\doi10.1103/PhysRevA.43.2046}{\bf 43}, 2046 (1991); M. Srednicki, Phys. Rev. E \href{\doi10.1103/PhysRevE.50.888}{\bf 50}, 888 (1994).

 \bibitem{QA}
J.-S.~Caux and F.H.L.~Essler, Phys. Rev. Lett. \href{http://dx.doi.org/10.1103/PhysRevLett.110.257203}{\bf 110}, 257203 (2013); M. Rigol, V. Dunjko, and M. Olshanii, Nature \href{http://www.nature.com/nature/journal/v452/n7189/full/nature06838.html}{\bf 452}, 854 (2008).


\bibitem{ProsenQLSymmetric} E. Ilievski, M. Medenjak \emph{et al}, J. Stat. Mech. (2016) \href{\doi10.1088/1742-5468/2016/06/064008}{064008} ; E. Ilievski, M. Medenjak, and T. Prosen, Phys. Rev. Lett. \href{http://dx.doi.org/10.1103/PhysRevLett.115.120601}{\bf 115}, 120601 (2015).


\bibitem{supp_matt_representation}
With respect to the parameter $u$ introduced in \cite{PereiraDrude}, we find more convenient to 
use the parameter $\alpha$, defined by $u = q^{s - \alpha}$ and focus on the neighborhood of $\alpha = 0$.

\bibitem{supp_matt}
See Supplementary Materials at [url].




\bibitem{CAD:hydro}
O. A. Castro-Alvaredo, B. Doyon, and T. Yoshimura, Phys. Rev. X \href{https://doi.org/10.1103/PhysRevX.6.041065}{\bf 6}, 041065 (2016). 

\bibitem{Prosen_vanishing_XXX}
J. M. P. Carmelo, T. Prosen, Nuclear Physics B \href{http://dx.doi.org/10.1016/j.nuclphysb.2016.10.021}{\bf 914}, 62-98 (2016).

%

\bibitem{Prosen_flux_quench}
M. Mierzejewski, P. Prelovsek, T. Prosen Phys. Rev. Lett. \href{10.1103/PhysRevLett.113.020602}{\bf 113}, 020602 (2014).

\bibitem{flux_quench}
O. N. Yuya,  G. Misguich and  M. Oshikawa, Phys. Rev. B \textbf{93}, 174310 (2016);
A. De Luca, Phys. Rev. B 90, 081403 (2014); 


\bibitem{Chern_Ins}
M. D. Caio, N. R. Cooper, and M. J. Bhaseen, Phys. Rev. B \href{10.1103/PhysRevB.94.155104}{\bf 94}, 155104.

\bibitem{twotemperatures}
D. Bernard and B. Doyon, 
  J. Phys. A: Math. Theor. \href{http://dx.doi.org/10.1088/1751-8113/45/36/362001}{\bf 45}, 362001 (2012);
A. De Luca, J. Viti \emph{et al}, 
 Phys. Rev. B \href{http://dx.doi.org/10.1103/PhysRevB.88.134301}{\bf 88}, 134301 (2013); C. Karrasch, R. Ilan, and J. E. Moore, 
Phys. Rev. B \href{http://dx.doi.org/10.1103/PhysRevB.88.195129}{\bf 88}, 195129 (2013);
M. Mintchev and P. Sorba, 
 J. Phys. A: Math. Theor. \href{http://dx.doi.org/10.1088/1751-8113/46/9/095006}{\bf 46}, 095006 (2013); A. De Luca, J. Viti \emph{et al}, Phys. Rev. B \href{https://doi.org/10.1103/PhysRevB.90.161101}{\bf 90}, 161101(R)  (2014);
B. Doyon, M. Hoogeveen, and D. Bernard, 
 J. Stat. Mech. (2014) \href{http://dx.doi.org/10.1088/1742-5468/2014/03/P03002}{P03002}; V. Eisler, Z. Zimboras, New J. Phys. \href{\doi10.1088/1367-2630/16/12/123020}{\bf 16}, 123020 (2014);
M. Collura and D. Karevski, Phys. Rev. B \href{http://dx.doi.org/10.1103/PhysRevB.89.214308}{89} 214308 (2014);
M. Collura and G. Martelloni, J. Stat. Mech. (2014) \href{http://dx.doi.org/10.1088/1742-5468/2014/08/P08006}{P08006}; 
A. De Luca, G. Martelloni, and J. Viti, 
 Phys. Rev. A \href{http://dx.doi.org/10.1103/PhysRevA.91.021603}{\bf 91}, 021603(R) (2015);
B. Doyon, A. Lucas \emph{et al}, 
  J. Phys. A: Math. Theor. \href{http://dx.doi.org/10.1088/1751-8113/48/9/095002}{48} 095002 (2015);
B. Doyon, 
  Nucl. Phys. B \href{\doi10.1016/j.nuclphysb.2015.01.007}{\bf 892}, 190 (2015); 
  J. Dubail, J.-M. Stephan, J. Viti and P. Calabrese,  \href{https://arxiv.org/abs/1606.04401}{arXiv:1606.04401}  (2016);
  D. Bernard, B. Doyon, \href{https://arxiv.org/abs/1612.05956}{arXiv:1612.05956} (2016).



\bibitem{bonnes14}
L. Bonnes, F.H.L. Essler and A. M. L\"auchli, 
Phys. Rev. Lett. \href{http://dx.doi.org/10.1103/PhysRevLett.113.187203}{\bf 113}, 187203 (2014).


\bibitem{sierrabook} C. G\'omez, M. Ruiz-Altaba, and G. Sierra. Quantum groups in two-dimensional physics. Cambridge University Press, 2005.
%




%
%
%
%
%
%
\end{thebibliography}
\end{document}